\documentclass[iop,revtex4]{emulateapj}
\usepackage{lscape}
\pdfoutput=1

\shorttitle{The VMC Survey. XIX. Classical Cepheids in the Small Magellanic Cloud}
\shortauthors{Ripepi et al.}

\begin{document}


\title{The VMC Survey. XIX. Classical Cepheids in the Small Magellanic Cloud}


\author{V. Ripepi\altaffilmark{1}}
\affil{INAF-Osservatorio Astronomico di Capodimonte, via Moiariello 16, 80131, Naples, Italy}
\email{ripepi@oacn.inaf.it}

\author{M. Marconi\altaffilmark{1}}
\affil{INAF-Osservatorio Astronomico di Capodimonte, via Moiariello 16, 80131, Naples, Italy}

\author{M. I. Moretti\altaffilmark{2}}
\affil{IAASARS, National Observatory of Athens, 15236 Penteli, Greece}

\author{G. Clementini\altaffilmark{3}}
\affil{INAF-Osservatorio Astronomico di Bologna, via Ranzani , 40127,  Bologna, Italy }

\author{M.-R. L. Cioni\altaffilmark{4,5,6}}
\affil{
 Universit\"{a}t Potsdam, Institut f\"{u}r Physik und Astronomie, Karl-Liebknecht-Str. 24/25, 14476 Potsdam, Germany \\
 Leibniz-Institut f\"{u}r Astrophysik Potsdam, An der Sternwarte 16, 14482 Potsdam Germany \\
 Centre for Astrophysics Research, School of Physics, Astronomy and Mathematics, University of Hertfordshire, College Lane, Hatfield AL10 9AB, UK
}

\author{R. de Grijs\altaffilmark{7,8}}
\affil{Kavli Institute for Astronomy \& Astrophysics and Department of
Astronomy, Peking University, Yi He Yuan Lu 5, Hai Dian District,
Beijing 100871, China \\
International Space Science Institute--Beijing, 1 Nanertiao, Zhongguancun, Hai Dian District, Beijing 100190, China}

\author{J. P. Emerson\altaffilmark{9}}
\affil{School of Physics \& Astronomy, Queen Mary 
University of London, Mile End Road, London E1 4NS, United Kingdom}

\author{M. A. T. Groenewegen\altaffilmark{10}}
\affil{ Koninklijke Sterrenwacht van Belgi\"e, Ringlaan 3, 1180, Brussel, Belgium}

\author{V. D. Ivanov\altaffilmark{11}}
\affil{European Southern Observatory, Karl-Schwarzschild-Strasse 2, 85748 Garching bei M\"unchen, Germany}


\author{A. E. Piatti\altaffilmark{12,13}}
\affil{Observatorio Astron\'omico, Universidad Nacional de C\'ordoba, Laprida 854, 5000, C\'ordoba, Argentina\\
Consejo Nacional de Investigaciones Cient\'ificas y T\'ecnicas, Av. Rivadavia 1917, C1033AAJ, Buenos Aires, Argentina
}


\altaffiltext{1}{INAF-Osservatorio Astronomico di Capodimonte, via Moiariello 16, 80131, Naples, Italy}
\altaffiltext{2}{IAASARS, National Observatory of Athens, 15236 Penteli, Greece}
\altaffiltext{3}{INAF-Osservatorio Astronomico di Bologna, via  Ranzani, 1, 40127, Bologna, Italy}
\altaffiltext{4}{ Universit\"{a}t Potsdam, Institut f\"{u}r Physik und Astronomie, Karl-Liebknecht-Str. 24/25, 14476 Potsdam, Germany }
\altaffiltext{5}{ Leibnitz-Institut f\"{u}r Astrophysik Potsdam, An der Sternwarte 16, 14482 Potsdam Germany }
\altaffiltext{6}{ University of Hertfordshire, Physics Astronomy and
  Mathematics, College Lane, Hatfield AL10 9AB, United Kingdom}
\altaffiltext{7}{Kavli Institute for Astronomy \& Astrophysics and Department of
Astronomy, Peking University, Yi He Yuan Lu 5, Hai Dian District,
Beijing 100871, China}
\altaffiltext{8}{International Space Science Institute--Beijing, 1 Nanertiao, Zhongguancun, Hai Dian District, Beijing 100190, China}
\altaffiltext{9}{School of Physics \& Astronomy, Queen Mary 
University of London, Mile End Road, London E1 4NS, United Kingdom}
\altaffiltext{10}{ Koninklijke Sterrenwacht van Belgi\"e, Ringlaan 3, 1180, Brussel, Belgium}
\altaffiltext{11}{European Southern Observatory, Karl-Schwarzschild-Strasse 2, 85748 Garching bei M\"unchen, Germany}
\altaffiltext{12}{Observatorio Astron\'omico, Universidad Nacional de C\'ordoba, Laprida 854, 5000, C\'ordoba, Argentina}
\altaffiltext{13}{Consejo Nacional de Investigaciones Cient\'ificas y T\'ecnicas, Av. Rivadavia 1917, C1033AAJ, Buenos Aires, Argentina}



\begin{abstract}
 The {\it VISTA near-infrared $YJK_\mathrm{s}$ survey of the
    Magellanic System} (VMC) is collecting deep $K_\mathrm{s}$-band
  time-series photometry of pulsating variable stars hosted by the two
  Magellanic Clouds and their connecting Bridge. In this paper, we
present  $Y,\,J,\,K_\mathrm{s}$ light curves for a sample of 4172 Small Magellanic Cloud (SMC) Classical
  Cepheids (CCs). These data, complemented with literature $V$ values, allowed us to construct a variety of
  period–-luminosity ($PL$), period–-luminosity–-color ($PLC$), and
  period-–Wesenheit ($PW$) relationships, valid for Fundamental (F), 
 First Overtone (FO) and Second Overtone (SO) pulsators. The relations involving $V,\,J,\,K_\mathrm{s}$ bands
 are in agreement with their counterparts in the
  literature. As for the $Y$ band, to our knowledge we present the
  first CC $PL$, $PW$, and $PLC$ relations ever derived using this
  filter. We also present the first near--infrared $PL$, $PW$, and $PLC$
  relations for SO pulsators to date. We used $PW(V,K_\mathrm{s})$ to
  estimate the relative SMC--LMC distance and, in turn, the absolute
  distance to the SMC. For the former quantity we find a value of 
  $\Delta\mu=0.55\pm0.04$ mag, in rather good
  agreement with other evaluations based on CCs, but 
  significantly larger than the results obtained from older population
  II distance indicators. This discrepancy  might be
  due to the different geometric distributions of young and old
  tracers in both Clouds. As for the absolute distance to the SMC, our
  best estimates are $\mu_{\rm SMC}=19.01\pm0.05$ mag and $\mu_{\rm
    SMC}=19.04\pm0.06$ mag, based on two distance measurements
 to the LMC, which rely on accurate CC and eclipsing Cepheid
  binary data, respectively. 
\end{abstract}





\section{Introduction}

The Magellanic Clouds (MCs) are fundamental touchstones in the context
of stellar populations and galactic evolution studies \citep[see,
  e.g.,][]{Harris2004,Harris2009,Ripepi2014a}. Indeed, they are
relatively close \citep[$D_{\sun}\sim50-60$
  kpc][]{Westerlund1997,Udalski1999}, they are rich in stars of
different ages, and their morphologies have been significantly
affected by their dynamical interaction. In effect, there are clear
signatures that the Small Magellanic Cloud (SMC), a gas-rich, dwarf
irregular galaxy, is interacting with its neighbours, the Large
Magellanic Cloud (LMC) and the Milky Way (MW). In particular, the MCs
are connected by a Bridge dominated by H{\sc i} gas but which also
contains a significant stellar content
\citep[e.g.][]{Irwin1985,Harris2007}. Like the Magellanic Stream, the
Bridge may be the signature of the MCs' mutual gravitational effects
and/or the impact of the MW \citep[e.g.][]{Putman1998,Hammer2015}.  In
addition, the SMC Wing, the part of the SMC main body extending
asymmetrically towards the LMC \citep{Shapley1940}, could be the
result of tidal interaction(s). Moreover, the bar of the SMC, traced
by the galaxy's young populations, appears highly asymmetric and
elongated, with its northeastern portion closer to us than its
southwestern part \citep[e.g.][]{Welch1987,Haschke2012,Rubele2015,Scowcroft2016}. In general,
the morphology of the SMC appears to depend on the age of the stellar
population used as a probe \citep[see, e.g.][and references
  therein]{Cioni2000a,Zaritsky2000,Dobbie2014,Deb2015}. The study of
the structure of the SMC is further complicated by the presence of a
considerable line-of-sight depth variation in the galaxy. Despite
several studies, it appears that the precise extent of the
line-of-sight depth and the three-dimensional (3D) geometry of the SMC
are still rather uncertain \citep[see, e.g.][for a
  review]{degrijs2014}. In fact, a comparison of the results in the
recent literature adopting different methods, namely horizontal-branch
stars, RR Lyrae and/or Classical Cepheid (CC) variables, red-clump
stars, full star-formation-history (SFH) reconstruction, star
clusters, etc., showed good qualitative agreement, but significant
discrepancies in the quantitive description of the geometry of the SMC
remain \citep[see, e.g.][and references
  therein]{Hatzidimitriou1989,Stanimirovic2004,Glatt2008,Nidever2013,Deb2015,Subramanian2015,Rubele2015}.

CC variables are at the base of the absolute calibration of the
extragalactic distance scale \citep[see, e.g.][and references
  therein]{Freedman2001,Marconi2005,Riess2011,Fiorentino2013} through
their well known Period--Luminosity ($PL$), Period--Luminosity--Color
($PLC$), and Period--Wesenheit ($PW$) relationships.

The CC $PL$ relations have been demonstrated, by several authors, to
show a nonnegligible dependence on both metallicity \citep[see,
  e.g.][and references therein]{Caputo2000,Romaniello2008,Bono2010}
and helium content
\citep[see][]{Fiorentino2002,Marconi2005,Carini2014}, and to exhibit a
nonlinear behavior towards the longest periods \citep[see, e.g.][and
  references therein]{Caputo2000,Ngeow2008,Marconi2009}. Both effects,
combined with the intrinsic dispersion due to the finite width of the
instability strip, are significantly reduced at near-infrared (NIR)
wavelengths \citep[][]{Bono1999,Caputo2000,Marconi2005,Marconi2010}.


The $PLC$ relations hold for each individual pulsator, since they
result from the combination of the period--density, the
Stefan--Boltzmann, and the Mass--Luminosity relations \citep[see,
  e.g.][for details]{Bono1999}, but they are affected by reddening and
metallicity uncertainties. On the other hand, the $PW$ relations are
reddening-free by definition \citep[e.g.][]{Madore1982, Caputo2000}
and include a color term that accounts at least in part for the finite
width of the instability strip. Moreover, they are less dependent on
chemical composition than the $PL$ relations. Furthermore, pulsation
amplitudes are much smaller in the NIR than in the optical bands, and
thus accurate mean magnitudes can be derived from a smaller number of
phase points along the pulsation cycle with respect to the optical
bands.

The {\it VISTA\footnote{Visible and Infrared Survey Telescope for
    Astronomy} near-infrared $YJK_\mathrm{s}$ survey of the Magellanic
  Clouds system} \citep[VMC; ][]{Cioni2011} aims at obtaining deep NIR
photometric data in the $Y$, $J$, and $K_\mathrm{s}$ filters over a
wide area covering the entire Magellanic system. VMC is a European
Southern Observatory (ESO) public survey that is carried out with
VIRCAM (VISTA InfraRed Camera) \citep{Dalton2006} on the ESO/VISTA
telescope \citep{Emerson2006}. The main goals of the survey are to
reconstruct the SFH and its spatial variation, as well as infer an
accurate 3D map of the entire Magellanic system. The properties of
pulsating stars observed by the VMC in the LMC and used as tracers of
three different stellar populations, namely CCs (younger than a few
hundred Myr), RR Lyrae and Type II Cepheid stars (older than 9--10
Gyr), and Anomalous Cepheids (traditionally associated with an
intermediate-age population of a few Gyr\footnote{However the
  possibility that they are old stars that underwent collisional or
  binary mergers cannot be excluded \citep[see, e.g.][and references
    therein]{Marconi04}}), have been discussed in recent papers by our
team
\citep{Ripepi2012a,Ripepi2012b,Moretti2014,Ripepi2014b,Muraveva2015,Ripepi2015}.
In these papers, we provided relevant results on the calibration of
the distance scale for all these important standard candles.

The scope of this paper is to present the results for the CCs in the
SMC after four years of VMC observations. The SMC is known to host
more than 4500 CCs, according to the OGLE\,III \citep{Soszynski2010}
and EROS\,2 \citep{Tisserand2007,Kim2014} surveys. The large number
of these pulsators, combined with their characteristic narrow
intrinsic $PL$, $PLC$, and $PW$ relationships in the NIR, make them
perfect tracers to unveil the complex structure of the SMC. Indeed, as
outlined above, the use of NIR relations has several advantages with
respect to the optical bands. Thus, the data presented in this paper
will allow us to study in more detail compared with past studies the
3D geometry of the galaxy. The results of that analysis will be
presented in a forthcoming paper.

This work is organized as follows. Sections 2 and 3 present the
observations and the technique used to fit the CC light curves,
respectively. Section 4 shows the color--magnitude diagrams and
peak-to-peak amplitudes; in Section 5 we illustrate the $PL$, $PLC$,
and $PW$ relationships obtained for the SMC CCs and the associated
results; a brief final Section 6 summarizes the paper.

\section{SMC Classical Cepheids in the VMC survey}

As referred to above, the two survey projects that identified CCs in
the SMC are OGLE\,III \citet[][]{Soszynski2010} and EROS\,2
\citep{Tisserand2007}. The areas covered by the two surveys overlap
almost completely, although OGLE\,III extends more towards the East,
whereas EROS\,2 covers a small corner in the North-West where
OGLE\,III data is not available \citep[see Fig. 4 in][for a
  comparison]{Moretti2014}.

In more detail, \citet[][]{Soszynski2010} reported the identification,
the $V,I$ light curves, and the main characteristics (periods, mean
magnitudes, etc.) of 4630 CCs in the SMC. The EROS\,2 collaboration
provided us with a list of CC candidates that was analyzed as
described in \citet{Moretti2014} to reject contaminating binaries,
resulting in 151 CC candidates. Among these objects, only about 20
were located outside the area investigated by OGLE\,III. A quick
comparison of the $PW$ in the $V,I$ bands\footnote{The Wesenheit
  magnitude in this case is defined as $W(V,I)=V-2.54(V-I)$. Note that
  EROS\,2 observations were carried out using custom $B_{\rm EROS\,2}$
  and $R_{\rm EROS\,2}$ filters that can be approximately converted to
  the Johnson $V,I$ bands using the transformation provided by
  \citet{Tisserand2007}.} revealed that the EROS\,2 CC candidates were
severely contaminated by other types of variables (typically Type II
Cepheids or Anomalous Cepheids) or by other unknown objects. To avoid
including spurious objects in our sample, we decided to use only
OGLE\,III data in the area covered by this survey, and to consider
only the $\sim20$ EROS\,2 CC candidates in the (small) area covered by
this survey but not by OGLE\,III. After removing from this small
sample those objects that were found to lie very far from the
OGLE\,III $PW$, we ended up with 13 bona fide CC candidates in the
EROS\,2-only field.

\begin{figure}
\epsscale{1.0}
\plotone{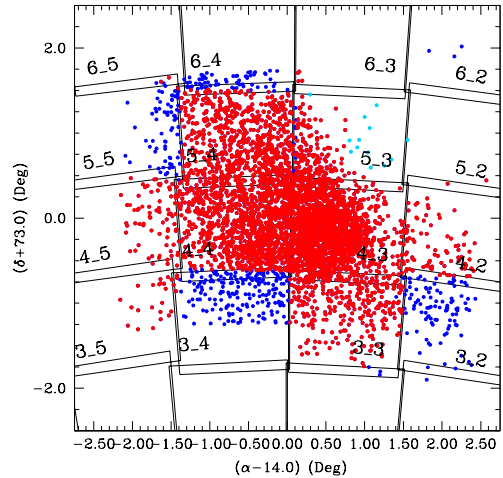}
\caption{Map of the CCs in the SMC. Red and blue filled 
  circles represent pulsators present in the OGLE\,III catalog and 
  indicated whether or not they have been observed by the VMC Survey,
  respectively. Light blue symbols show the 13 Cepheids identified on the 
  basis of the EROS\,2 data 
  \citep[][]{Tisserand2007,Moretti2014}.\label{fig1}}
\end{figure}

In this paper we present results for the CCs included in 11 tiles
(each tile is 1.5 deg$^2$ on the sky) completely or nearly completely
observed, processed, and catalogued by the VMC survey as of 2015 March
9 (including observations obtained until 2014 September), namely the
tiles SMC 3\_3, 3\_5, 4\_2, 4\_3, 4\_4, 4\_5, 5\_2, 5\_3, 5\_4, 6\_3,
and 6\_5. Figure~\ref{fig1} shows the spatial extent of the VMC tiles
across the SMC body. The completed tiles do not cover the entire area
surveyed by OGLE\,III. However, given the high concentration of CCs in
the central body of the SMC, the number of pulsators included in the
completed VMC tiles is about 90\% of the total OGLE\,III
sample. Table~\ref{ncep1} lists the coordinates of the quoted tiles,
as well as the number of CCs included in each tile.

In total, we were able to study 4159 objects of the 4630 OGLE\,III
sample. To this number we have to add the 13 CCs from the EROS\,2
data, leaving us with a final sample of 4172 CCs. The classification
of the investigated pulsators in terms of Fundamental (F), First
Overtone (FO), Second Overtone (SO), and mixed modes (F/FO, FO/SO,
F/FO/SO, FO/SO/TO, where TO stands for Third Overtone) is shown in
Table~\ref{ncep2}.

\begin{figure}
\plotone{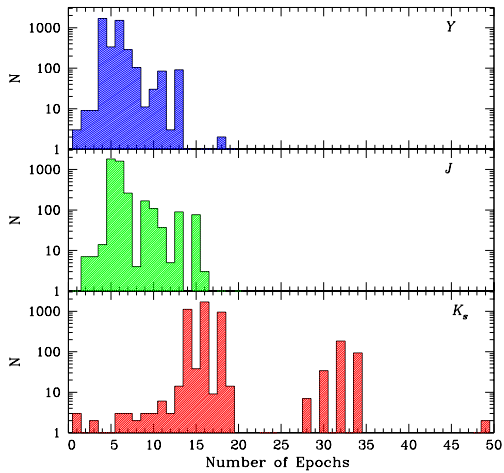}
\caption{Histogram of the numbers of epochs in each photometric band. 
 \label{fig2}}
\end{figure}

\begin{figure*}
\epsscale{0.9}
\plottwo{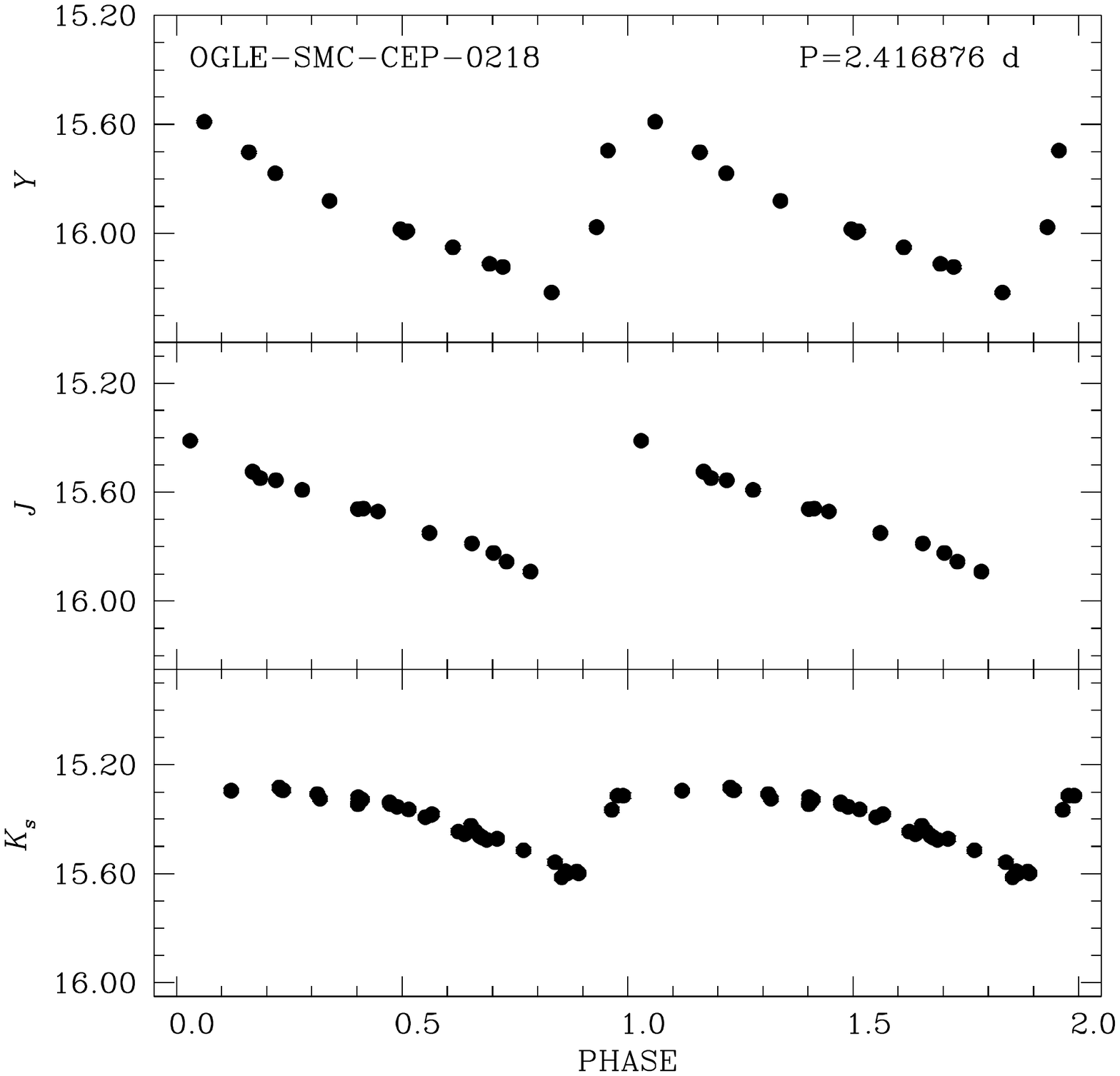}{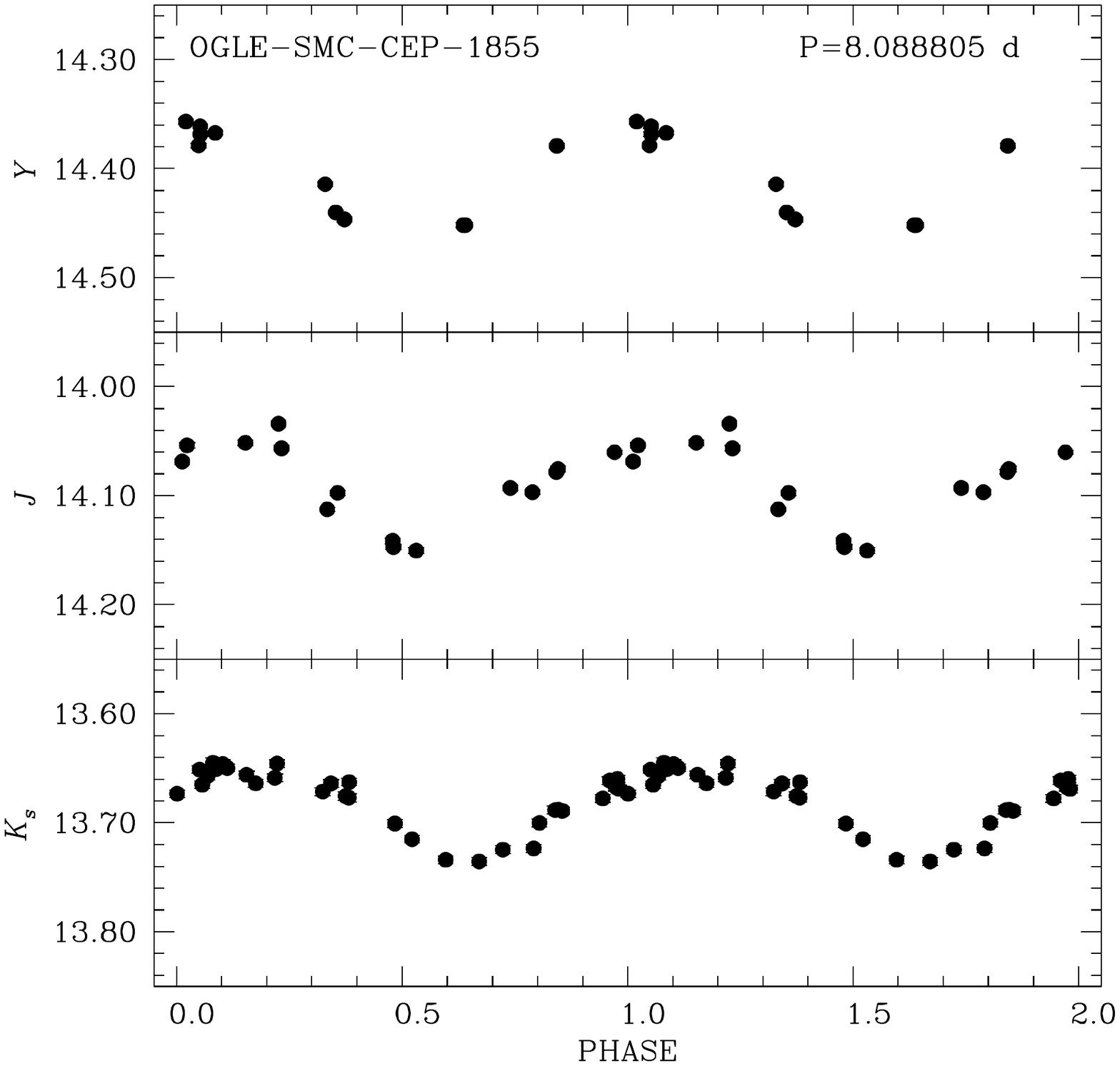}
\caption{$Y\,J\,K_\mathrm{s}$ light curves for the labelled 
  Cepheids. The errors are of similar size or smaller than the size of 
  the symbols. Note the quality of the data and the change in 
  amplitude and shape of the light curve going from the $Y$ to the 
  $K_\mathrm{s}$ bands.\label{fig3}}
\end{figure*}

A general description of the observing strategy of the VMC survey can
be found in \citet{Cioni2011}. The procedures adopted to study the
variable stars were discussed in detail by
\citet{Ripepi2012a,Ripepi2012b,Moretti2014,Ripepi2014b,Ripepi2015}.
However, it is worth recalling that the VMC $K_\mathrm{s}$-band
time-series observations were scheduled to span 13 separate epochs
distributed over several consecutive months. This observing strategy
permits one to achieve well-sampled light curves for different types
of variable stars, including RR Lyrae variables and Cepheids of all
types. As for the $Y$ and $J$ bands, the nominal number of epochs is
four (two of these epochs are obtained with half exposure time) and
may be acquired during the same night given that monitoring 
in these filters was not planned. However, a few additional
epochs are usually available for each tile (especially in the
$K_\mathrm{s}$-band), 
because some observing
blocks (OBs) were executed outside of our specifications (typically
for seeing values exceeding 0.8--1.0 arcsec), but the data were still
useful since the CCs are relatively bright. In addition, there is a
small overlap between the tiles. Consequently, the CCs present in
multiple tiles possess at least twice the scheduled number of
epochs. Given the high concentration of CCs in the contiguous tiles
SMC 4\_3, 4\_4, 5\_3, and 5\_4 (see Fig.~\ref{fig1}), we have more
than 320 CCs whose light curves contain more than $\sim$28 phase
points. This is also shown in the bottom panel of Fig.~\ref{fig2},
where the bimodal distribution of epochs in the $K_\mathrm{s}$ band is
clear. From the figure, note that there are a few dozen stars with
fewer than 13 epochs in $K_\mathrm{s}$. This can happen when the
sources are located in underexposed areas and/or are affected by
bright neighbours or bad pixels. We were still able to analyze these stars
thanks to our template-fitting procedure (see Section~\ref{template}).

The same considerations hold for the $Y$ and $J$ bands, whose epoch
distributions are shown in the top and middle panels of
Fig.~\ref{fig2}, respectively. In this case, the number of CCs with
more than 10 epochs is 213 and 321 in the $Y$ and $J$ bands,
respectively. Similarly, the number of CCs with more than five epochs
is 2121 and 2343 in $Y$ and $J$, respectively.
 
The VMC data were processed with the pipeline \citep{Irwin2004} of the
VISTA Data Flow System \citep[VDFS,][]{Emerson2004} and the photometry
is in the VISTA photometric system (Vegamag=0). The time-series data
analyzed in this work were downloaded from the VISTA Science
Archive\footnote{http://horus.roe.ac.uk/vsa/}
\citep[VSA,][]{Cross2012}.
Details about the data reduction can be found in the aforementioned
papers. However, we briefly recall that (i) the pipeline applies a
correction to the photometry of stars close to the saturation limit
\citep{Irwin2009}. This task is very useful, because long-period CCs
are very bright ($K_\mathrm{s} \sim 12$--13) mag. The time-series
photometry of these variables takes advantage of the VDFS capability
to deal with the images for saturation, although the corrections
applied do not always guarantee a full recovery of the data. (ii) The
VSA processing produces quality flags for each star that are valuable
to understand if the images have problems. This information is
important for the following analysis.

To obtain the $Y$, $J$, and $K_\mathrm{s}$ light curves, the OGLE\,III
(and EROS\,2) catalog(s) of CCs described above were cross-correlated
against the VMC catalog, taking all counterparts from the OGLE\,III
and EROS\,2 positions, regardless of their separation on the sky.
About 95\% of the objects have positions in agreement with those
measured by OGLE\,III and EROS\,2 within less than 0.1$''$. Among the
remaining 186 stars, 67 have a separation larger than 0.5$''$ and are
likely misidentifications. We will come back to these objects below.

The typical quality of the light curves obtained is illustrated in
Fig.~\ref{fig3} for two F pulsators with very different periods.
 VMC photometry for the 4172 stars analysed in this work is
  reported in Table~\ref{VMCPhot}. The complete version of the table
  is available online at the journal site.

\section{Template-fitting procedure}
\label{template}

Given the large number of light curves to analyze, it was convenient
to find an automatic way to process the data. Our aim is to obtain an
analytical or empirical model light curve that fits the observed
one. This model can subsequently be used to measure the mean magnitude
and the peak-to-peak amplitude for each variable. The usual way to
carry out such a task is to use truncated Fourier series, adding as
many harmonics as needed to obtain a good fit to the data \citep{Schaltenbrand1971}. However,
this kind of approach would not be useful in our case, because the
presence of significant gaps in the light curve would lead to strong
and unrealistic oscillations in the Fourier series.

Hence, we decided to use template light curves to fit the data.
 Following the pioneering work by \citet{Freedman1988}, templates 
to fit CC light curves based on only a few epochs in the NIR
bands have already been presented and used by
\citet{Soszynski2005,Inno2013,Inno2015}. The typical approach in these
studies consists of the following steps: (1) adopting templates
constructed based on well-sampled CC $J$,$H$, and $K_\mathrm{s}$ light
curves (Galactic or MC objects); (2) scaling the template amplitude
using fixed amplitude ratios (e.g., A($J$)/A($V$), with some
dependence on period); (3) adopting literature period and epoch of
maximum light to phase-match the template and the observed data. This
technique is valuable, since it allows one to obtain an estimate of
the average magnitude of a CC based on just one or two observed phase
points. At the same time, given the uncertainties on the
  amplitude ratios and on the ephemerides, these
  estimates can easily be affected by errors as large as 5\% (see also
  Sect. 3.3),  despite the quite low amplitudes of the light
curves in the NIR bands. 

Our approach is fundamentally different from that outlined above
\citep[e.g., by][]{Inno2013}. Indeed, the availability of an average
of $\sim$ 5.7, 6.3, and 16.7 phase points in $Y$, $J$, and
$K_\mathrm{s}$, respectively, allows us to safely rescale our
templates in amplitude and phase match them using our observations
directly. This procedure eliminates most of the uncertainties of the
``classical'' template method, because {\it we do not have to rely on
  any fixed amplitude ratio to scale the templates in amplitude, nor
  do we have to use the literature epoch of maximum as reference to
  phase match the template and the observed data}.

\begin{figure}
\epsscale{1.0}
\plotone{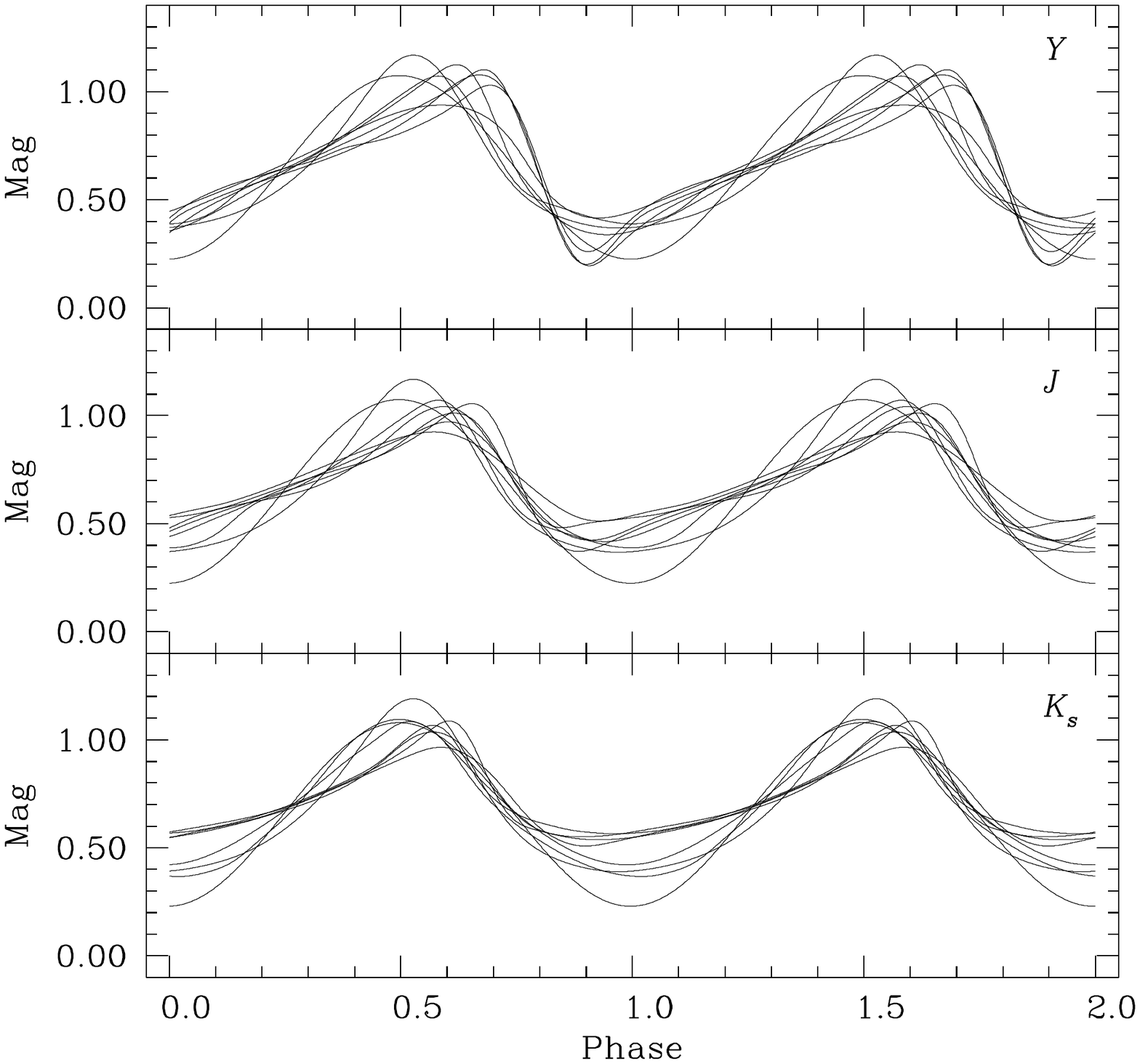}
\caption{Adopted templates in the $Y\,J\,K_\mathrm{s}$ bands. 
\label{allTemplate}}
\end{figure}


\begin{figure*}
\epsscale{1.0}
\plotone{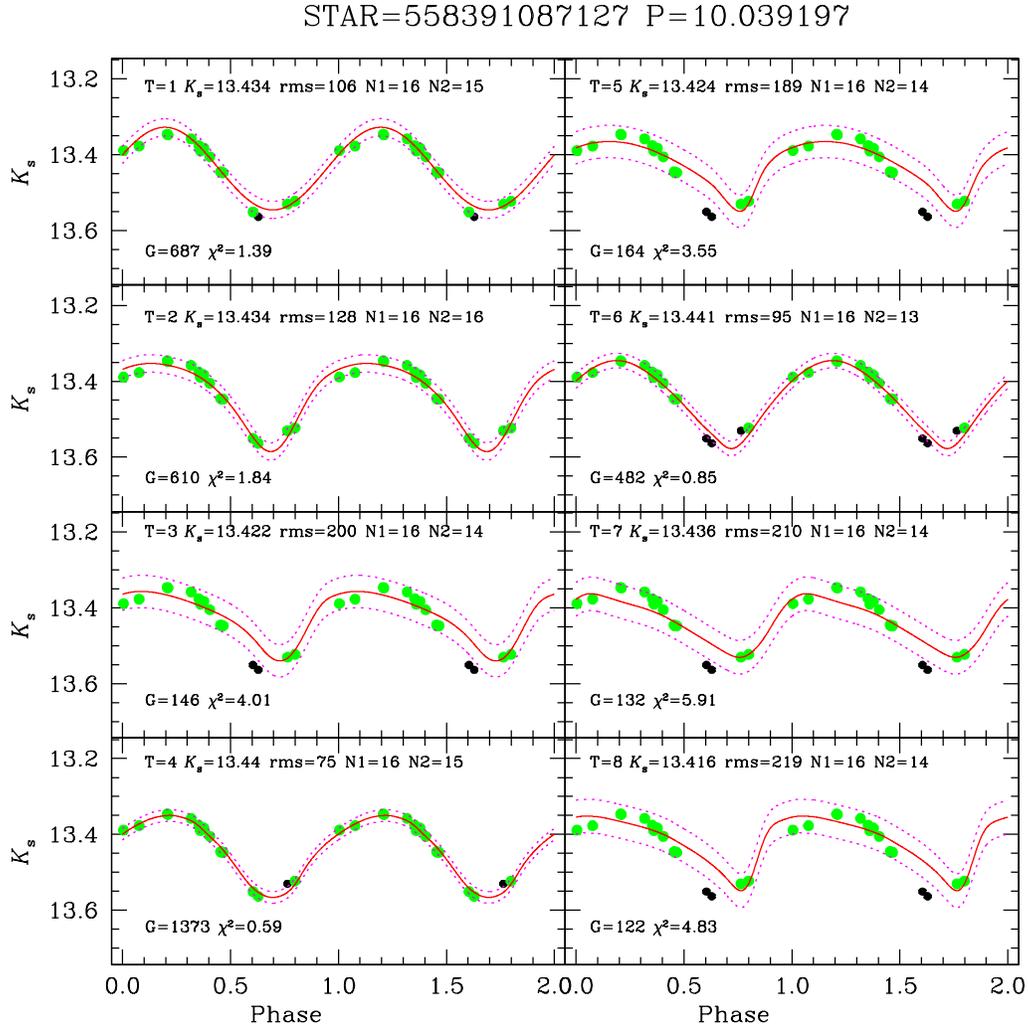}
\caption{Example of our template-fitting procedure in the 
  $K_\mathrm{s}$ band. Each of the eight panels shows the 
  $K_\mathrm{s}$ light curve (green and black filled circles show the 
  data points used and rejected in the fitting procedure,
  respectively). The solid lines are the template curves (labelled 
  with ``T'' in each panel, with increasing numbers from 1 to 8),
  properly scaled in amplitude and shifted in phase. The dashed lines 
  represent the $\pm 2 \sigma$ template curves: all data points beyond 
  these lines are marked in black and not included in the fitting 
  procedure. The other labels in each panel are: $K_\mathrm{s}$ = mean 
  magnitude of the curve needed to fit the data with the template;
  r.m.s. = root mean square of the fit residuals in mmag; $N1$ = total 
  number of data points in the light curve; $N2$ = number of data 
  points used in the fitting procedure; $G$ = goodness parameter (see 
  the text); $\chi^2$ = $\chi^2$ of the fit (see the text). In this 
  case, the best template is T4.  \label{template}}
\end{figure*}

\begin{figure}
\plotone{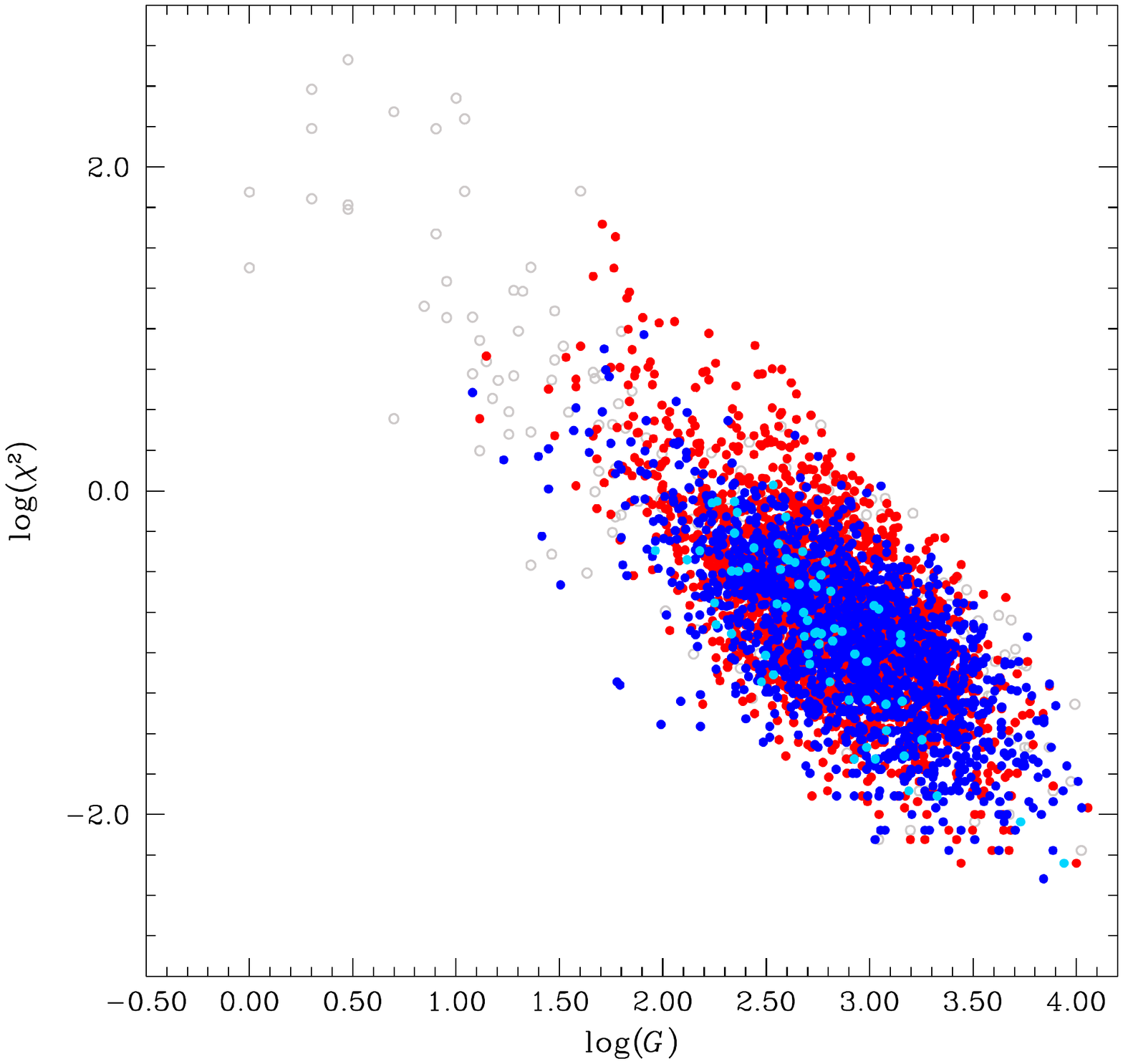}
\caption{Goodness versus $\chi^2$ in the $K_\mathrm{s}$ band for the 
  target CCs. F, FO, and SO pulsators are shown as red, blue, and 
  light blue filled circles, respectively. The gray open circles show the 
  objects excluded from the analysis on the basis of their location 
  near the $PW(K_\mathrm{s},V)$ relation (see 
  Section~\ref{relations}). 
\label{logG}}
\end{figure}

\subsection{Template construction}

The first step of our procedure was the construction of the
templates. To this end, we visually inspected a large number of light
curves, trying to select those with the most often recurring shapes,
and at the same time, those exhibiting precise light curves.
Particular care was devoted to covering a broad range of periods. This
search was rather simple in $K_\mathrm{s}$, since in this band we have
dozens of well-sampled and precise light curves for any
period. However, we had fewer choices in $Y$ and $J$, given the much
smaller number of well-sampled light curves in those filters.

At the end of this process, we concluded that a set of eight different
templates for each band could reproduce the variety of shapes
exhibited by the observed light curves.

Our templates were constructed as truncated Fourier series of the
form
\begin{equation}
m(\phi)=a_0+\sum\limits_{k=1}^N [a_k {\rm cos}(2 \pi k \phi +\Phi_k)],
\end{equation}
\noindent
where $m$ is the magnitude, $\phi$ are the phases of the template
light curves, $a_0$ is the zero point, which is zero by definition,
$N$ is the number of terms of the series, and $a_k$ and $\Phi_k$ are
the amplitudes and phases of each term of the series,
respectively. The first step consisted of fitting the selected
observed light curves with splines in order to have smooth, densely
sampled curves to be passed to the Fourier-series fitting
program. This was needed to avoid spurious oscillations in the
Fourier-series fit due to possible small gaps or undersampling at
maximum/minimum of the light curves. This way, we actually constructed
six of the eight different templates adopted for each filter. They are
listed in Table~\ref{paramFourier}, from T3 to T8. As for the two
remaining templates, T1 simply consists of a pure cosine function for
all filters (which is why the T1 template is not included in the
table), while T2 reproduces a smooth curve which can often be observed
in all of the $Y, J$, and $K_\mathrm{s}$ bands for a broad range of
periods (see T2 in Table~\ref{paramFourier}). The shapes of the eight
templates in all three filters are shown in Fig.~\ref{allTemplate}.

\subsection{Template fitting}

The template-fitting procedure includes the following steps: 

\begin{itemize} 

\item Scaling the templates with amplitude ratios, e.g.,
  A($K_\mathrm{s}$)/A($I$), where we take A($I$) from the
  OGLE\,III survey and the coefficients of these ratios from
  \citet{Soszynski2005}. Similarly, the template is phase-matched with
  the observations using the ephemerides from OGLE\,III. The purpose
  of this step is {\it only} to provide a first-guess average
  magnitude for the target star and, in turn, to remove the most
  obvious outliers. In practice, we simply estimate the template
  values at the phases of the observed point and calculate the average
  difference between observed and calculated values, which is the
  approximate mean magnitude of the star.

\item Recalculating the template by varying its initial phase to find
  the phase shift that minimizes (by means of a least-squares fit) the
  difference data--template. This step provides an improved average
  magnitude (of the order of a few hundredths of mag).

\item Recalculating the template by varying its amplitude to find the
  amplitude scaling that minimizes (by means of a least-squares fit)
  the difference data--template. This step provides a further
  improvement of the average magnitude (again, a few hundredths of mag).

\item Fine-tuning outlier removal (at 2$\sigma$ and 3$\sigma$ levels
  in $K_\mathrm{s}$ and in $Y, J$, respectively; the difference is
  because in $Y,J$ we have many fewer phase points than in
  $K_\mathrm{s}$ and, hence, we can simply remove obvious outliers)
  and final average-magnitude calculation (in intensity).

\end{itemize}

This procedure is applied to each template (in each band). Next, we
need a tool to choose the template that optimally represents the
data. After several trials and visual inspections of the resulting
fits, we devised two main useful diagnostics. The first is the usual
$\chi^2$ minimization, defined as
\begin{equation}
\chi^2\,=\,\frac{1}{N}\sum\limits_{\phi=0}^1 \left(\frac{m(\phi)_{\rm
      obs} - m(\phi)_{\rm temp}}{\sigma(m(\phi)_{\rm obs})}\right)^2,
\end{equation}
\noindent 
where $N$ is the total number of phase points, $m_{\rm obs}$ and
$m_{\rm temp}$ are the magnitudes of the observed and computed light
curves, respectively, and $\sigma(m_{\rm obs})$ is the magnitude
uncertainty per phase point.
 
\begin{figure}
\plotone{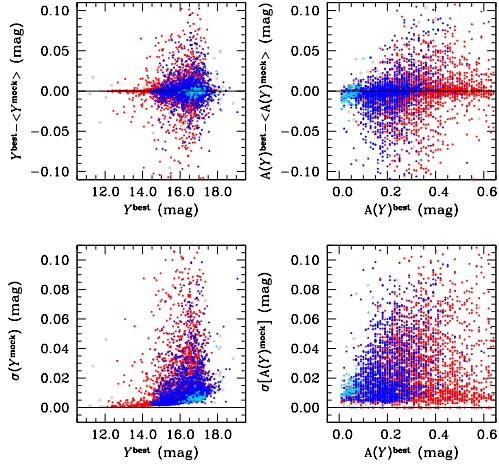}
\caption{Results from the Monte Carlo experiments in the $Y$ band. The 
  top left-hand panel shows, for each star, the difference between the 
  magnitude obtained with the best-fitting template ($Y^{\rm best}$) 
  applied to the real data and that resulting from averaging over the 
  100 Monte Carlo experiments ($\langle Y^{\rm mock}\rangle$).  The 
  bottom left-hand panel shows the r.m.s. of $Y^{\rm mock}$ as a 
  function of $Y^{\rm best}$. The top and bottom right-hand panels are 
  similar to those on the left but display the peak-to-peak amplitudes 
  instead of the magnitudes (in the $Y$ band). The color coding is the 
  same as in Fig.~\ref{logG}. \label{figMC_Y}}
\end{figure}

\begin{figure}
\plotone{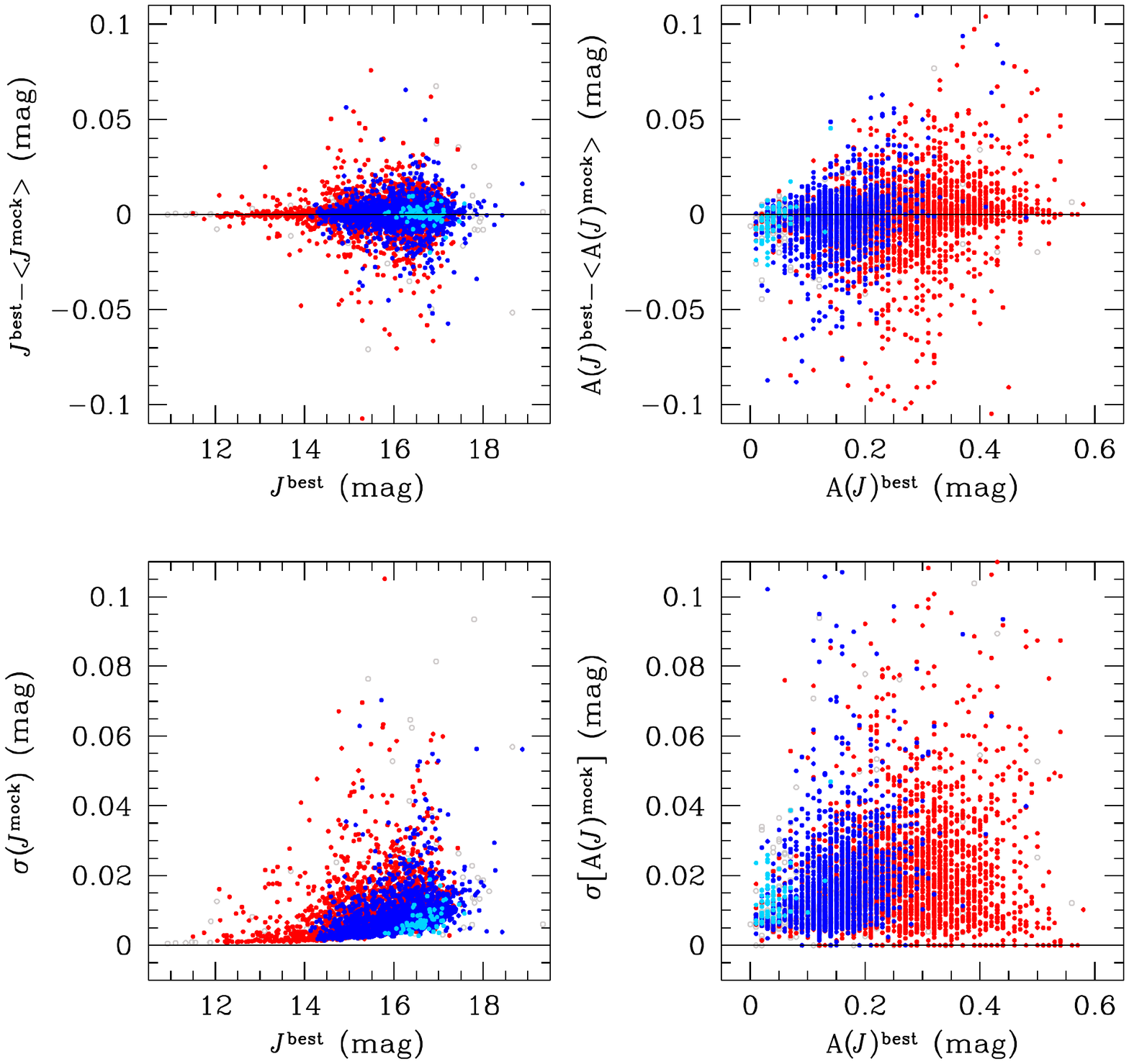}
\caption{As Fig.~\ref{figMC_Y} but for the $J$ band. 
\label{figMC_J}}
\end{figure}

\begin{figure}
\plotone{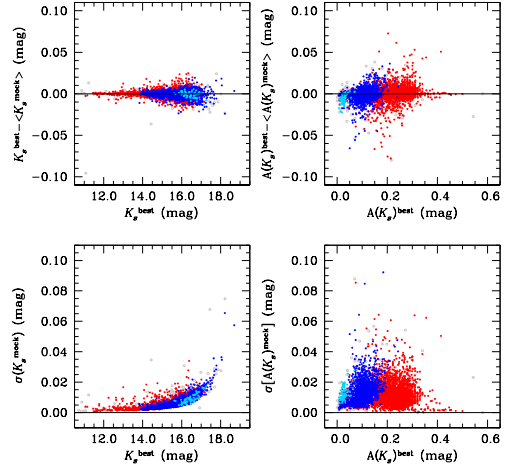}
\caption{As Fig.~\ref{figMC_Y} but for the $K_\mathrm{s}$
  band.  \label{figMC_K}}
\end{figure}

\begin{figure*}
\plotone{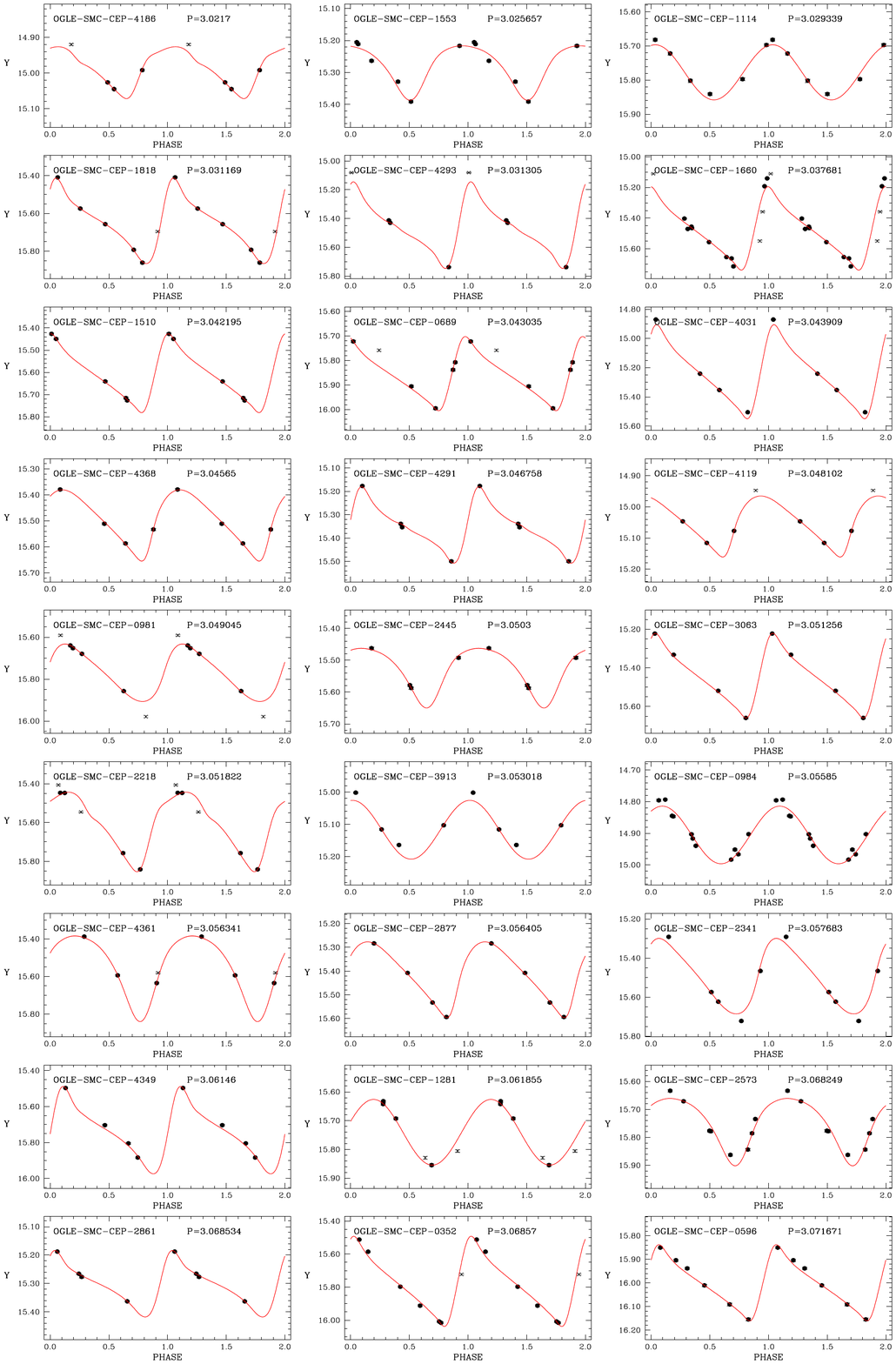}
\caption{Light curves in the $Y$ band with superimposed the best 
  template fit for a sample of 27 CCs analyzed in this paper. Filled 
  circles and crosses show the data points adopted and discarded 
  during the fitting procedure, respectively. The solid red line 
  represents the template fitted to the data. The OGLE\,III or EROS\,2 
  identification and the period of the variable are also reported.  
The complete figure, including the light curves for the full data 
  set of 4172 CCs, is published in its entirety in the  
electronic edition of the {\it Astrophysical Journal}.  A portion is  
shown here for guidance regarding its form and content.  
Note  that in the electronic version of the figure, the CC light curves are 
  shown in order of increasing period (see caption of 
  Table~\ref{tabResults} for details). 
\label{figY_pag119}}
\end{figure*}

\begin{figure*}
\plotone{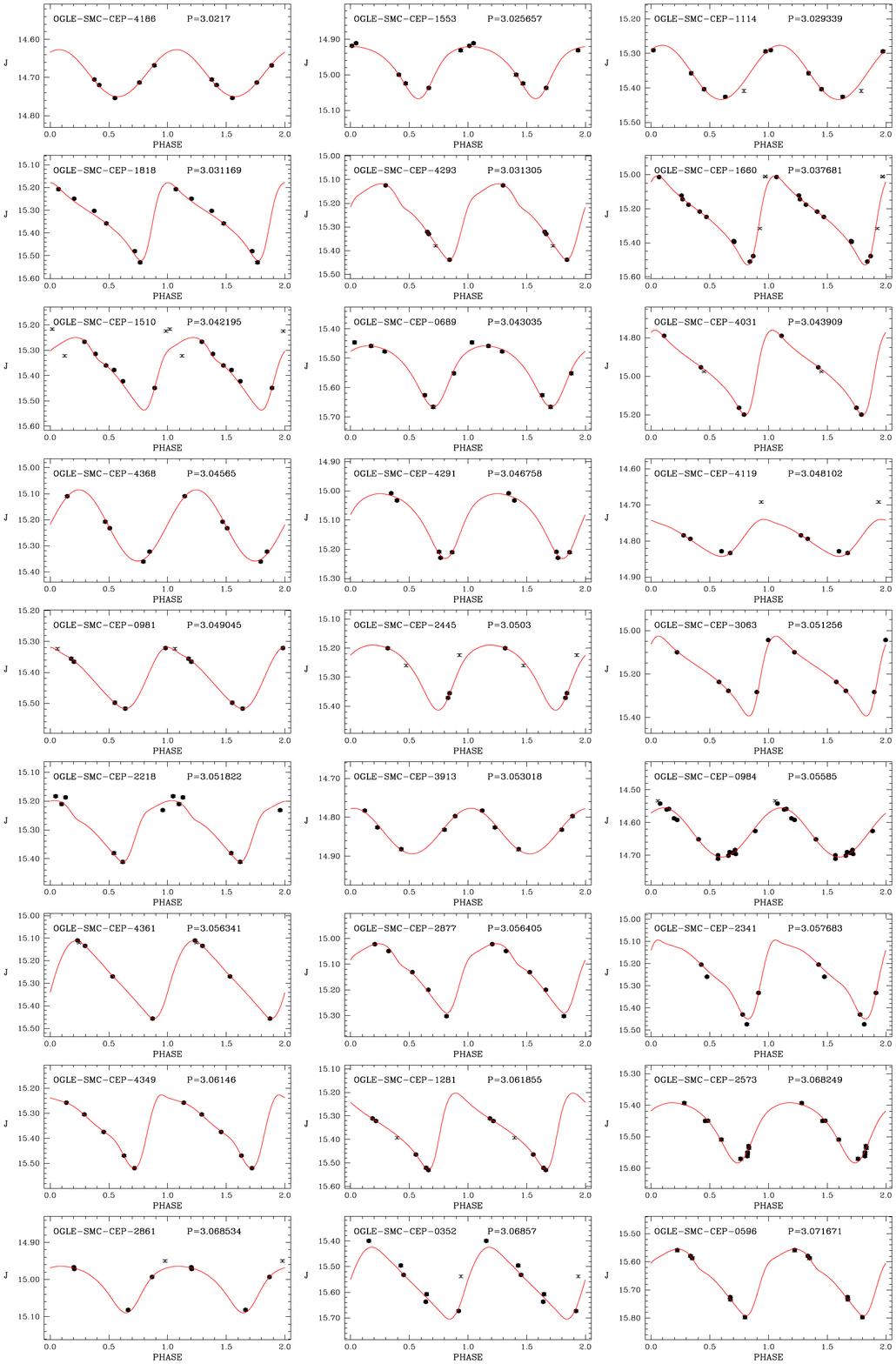}
\caption{As Fig~\ref{figY_pag119} but for the $J$ band. 
 \label{figJ_pag119}}
\end{figure*}

\begin{figure*}
\plotone{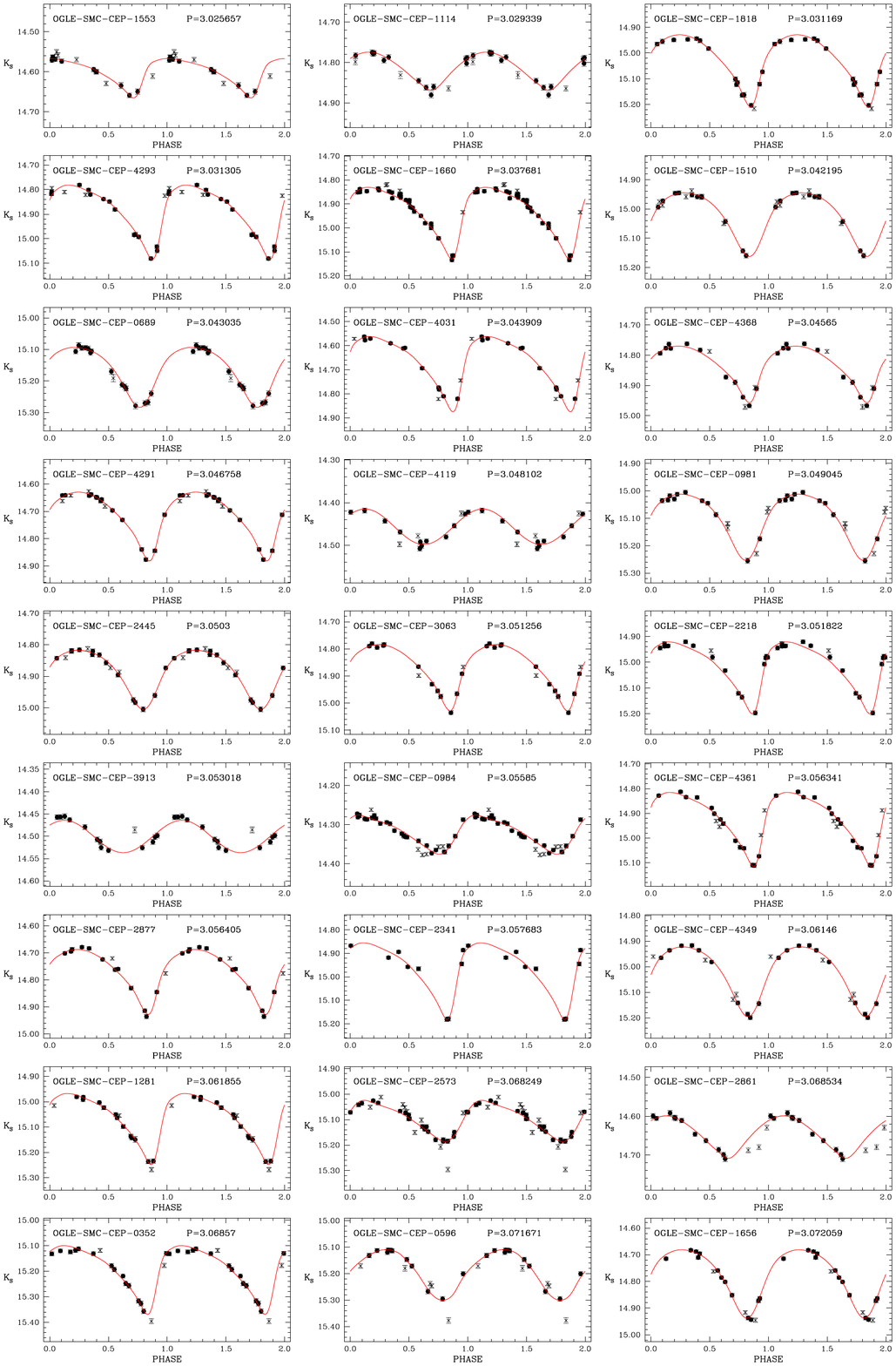}
\caption{As Fig~\ref{figY_pag119} but for the $K_\mathrm{s}$ band. 
 \label{figK_pag119}}
\end{figure*}

The second criterion was devised empirically to take into account the
fact that the smallest residuals can result from application of the
wrong template simply because the outlier-removal process is too
aggressive. Thus, we designed a {\it Goodness} (or {\it G}) parameter,
defined as
\begin{equation}
G\,=\,{\rm int}\left[\left(\frac{1}{\sigma}\right)^2\, \left(\frac{N_{\rm U}}{N_{\rm T}}\right)^4\,10^6\right],
\end{equation}
\noindent
where $\sigma$ is the r.m.s. of the fit and $N_{\rm U}$ and $N_{\rm
  T}$ are the numbers of phase points used in the fit and the total
number of phase points, respectively. By definition, the first factor
of $G$ tends to favor templates that give the smallest r.m.s. values,
while the second factor favors those that remove the lowest number of
outliers. The balance of these two features yields, in general, an
automatic decision about the best templates that is in agreement with
visual inspection of the fitting procedure. The value of $G$ can be
used not only to choose the best template, but also as a more general
indicator of the quality of the data and of the relative fitting
procedure. Indeed, in general high values of $G$ (in our case,
typically $G>100$) mean good data (and good fits), while lower values
usually indicate significant scatter in the light curves. Extremely
high values of $G$ (i.e., $G>10,000$) are also rather suspect because
non-variable stars, exhibiting completely flat light curves (which
happens, for example, when non-variable stars are considered owing to
a coordinate missmatch with OGLE\,III Cepheid data) are expected to
yield extremely high values of $G$. Not surprisingly, the $G$
parameter is anti-correlated with the corresponding $\chi^2$ value:
the higher $G$ is, the lower the $\chi^2$ becomes. An example of our
template-fitting procedure can be found in Fig.~\ref{template} (note
that in this case the best template is T4), while the anti-correlation
between $G$ and $\chi^2$ is shown in Fig.~\ref{logG}.

\subsection{Monte Carlo simulations}

As an additional check of the reliability of the template-fitting
procedure, and to estimate in a more quantitative way the precision
achieved, we decided to use extensive Monte Carlo simulations. In
practice, for each star (and for each filter), 100 different mock time
series were created on the basis of the observed data, to which
Gaussian noise was added with $\sigma$'s corresponding to the average
uncertainty on the phase points for the star of interest (different
$\sigma$'s were calculated for different filters). The
template-fitting procedure outlined above was hence repeated 100 times
and the average magnitude and r.m.s. were calculated. We then compared
these quantities with those calculated from the observed data. The
results of this exercise are summarized in Figs~\ref{figMC_Y},
\ref{figMC_J}, and \ref{figMC_K}. The top left-hand panels in each
figure show the difference between the magnitude obtained with the
best-fitting template ($Y^{\rm best},\,J^{\rm
  best},\,K_\mathrm{s}^{\rm best}$) applied to the actual data and
that resulting from averaging over the 100 mock light curves ($\langle
Y^{\rm mock}\rangle,\,\langle J^{\rm mock}\rangle,\,\langle
K_\mathrm{s}^{\rm mock}\rangle$). Similarly, the bottom left-hand
panels show the r.m.s. of $Y^{\rm mock},\,J^{\rm
  mock},\,K_\mathrm{s}^{\rm mock}$ as a function of $Y^{\rm
  best},\,J^{\rm best},\,K_\mathrm{s}^{\rm best}$. These figures
testify to the high precision reached in the $K_\mathrm{s}$ band,
where 84\% and 99\% of the stars have errors on the intensity-averaged
magnitudes of $\leq\,0.01$ mag and $\leq 0.02$ mag respectively. Only
1\% and 0.1\% of the CCs analyzed here have uncertainties $>\,0.02$
mag and $>\,0.05$ mag, respectively. The results are less favorable in
the $J$ band and even worse in $Y$. In fact, in these bands the
corresponding percentages drop to (68\%, 93\%, 7\%, and 0.8\%) and
(56\%, 78\%, 22\%, and 5.5\%) in the $J$ and $Y$ bands,
respectively. The worse results in $Y$ are mainly due to the fact that
(a) the peak-to-peak amplitude in this filter is significantly larger
than that in the $J$ band (consequently, it is more difficult to
estimate the average magnitudes from undersampled light curves) and
(b) we have, on average, fewer phase points in $Y$ than in $J$
($\sim$5.6 versus $\sim$6.3).

\begin{figure*}
\hbox{
\includegraphics[scale=.27]{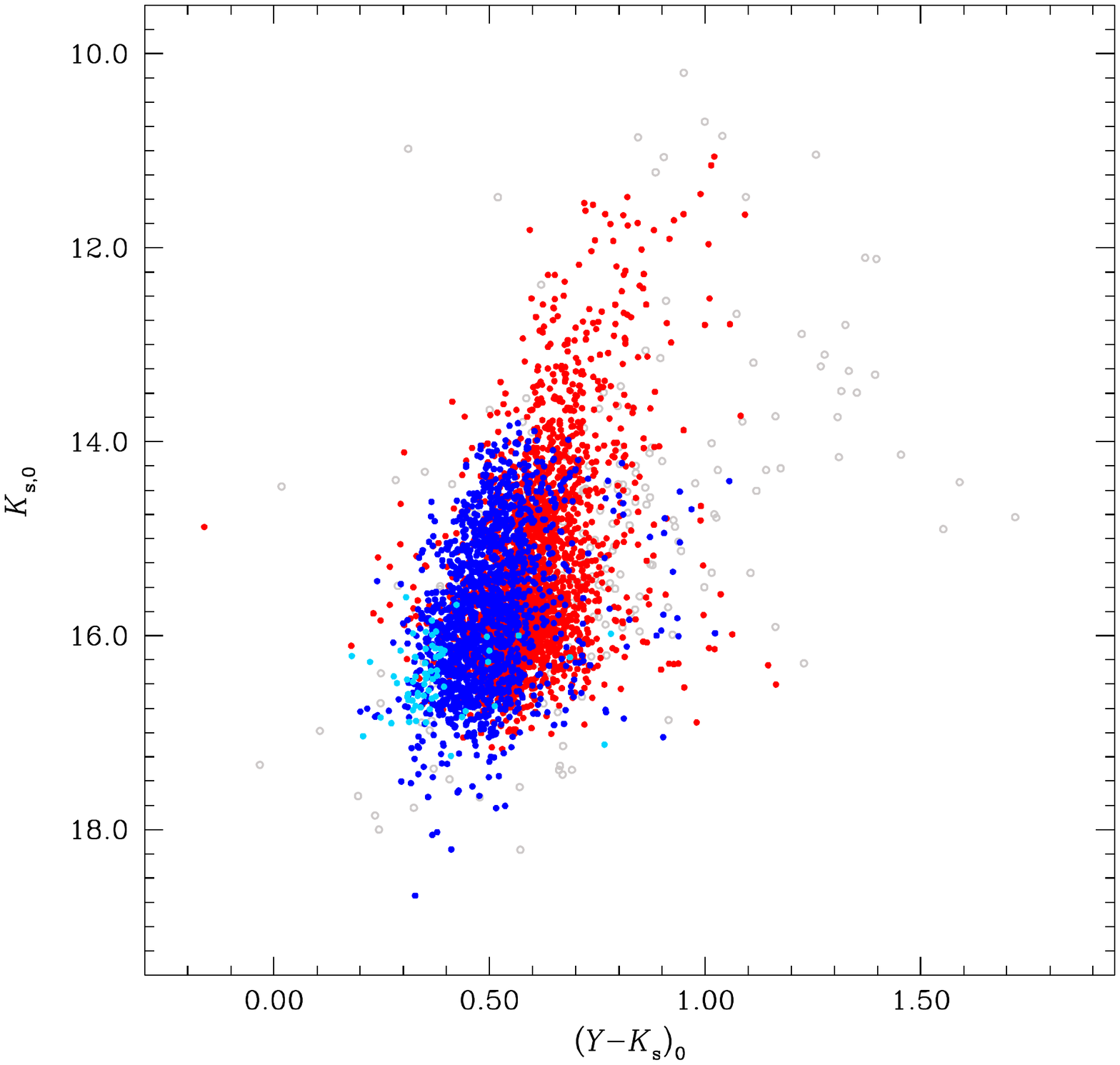}
\includegraphics[scale=.27]{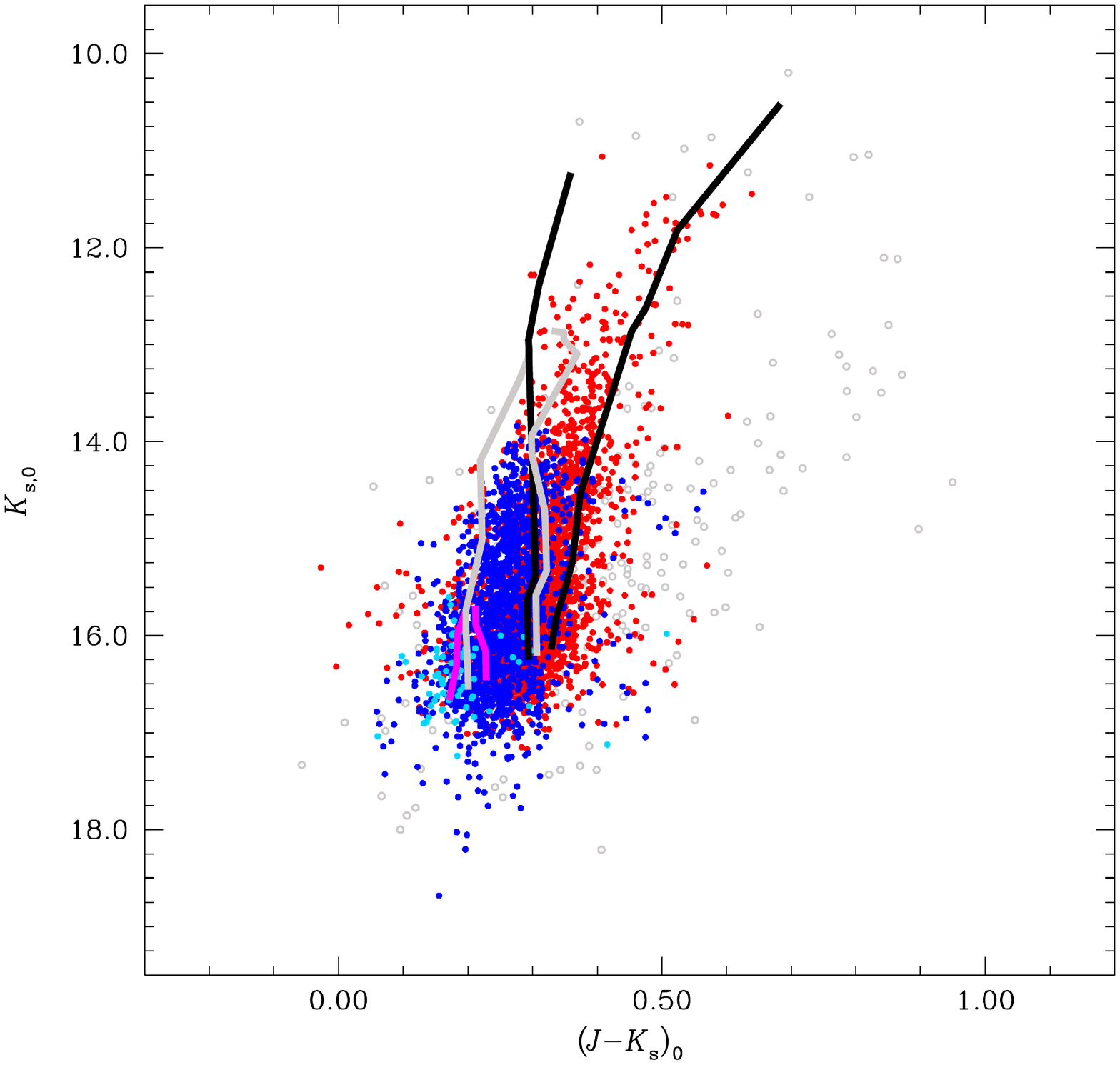}
\includegraphics[scale=.27]{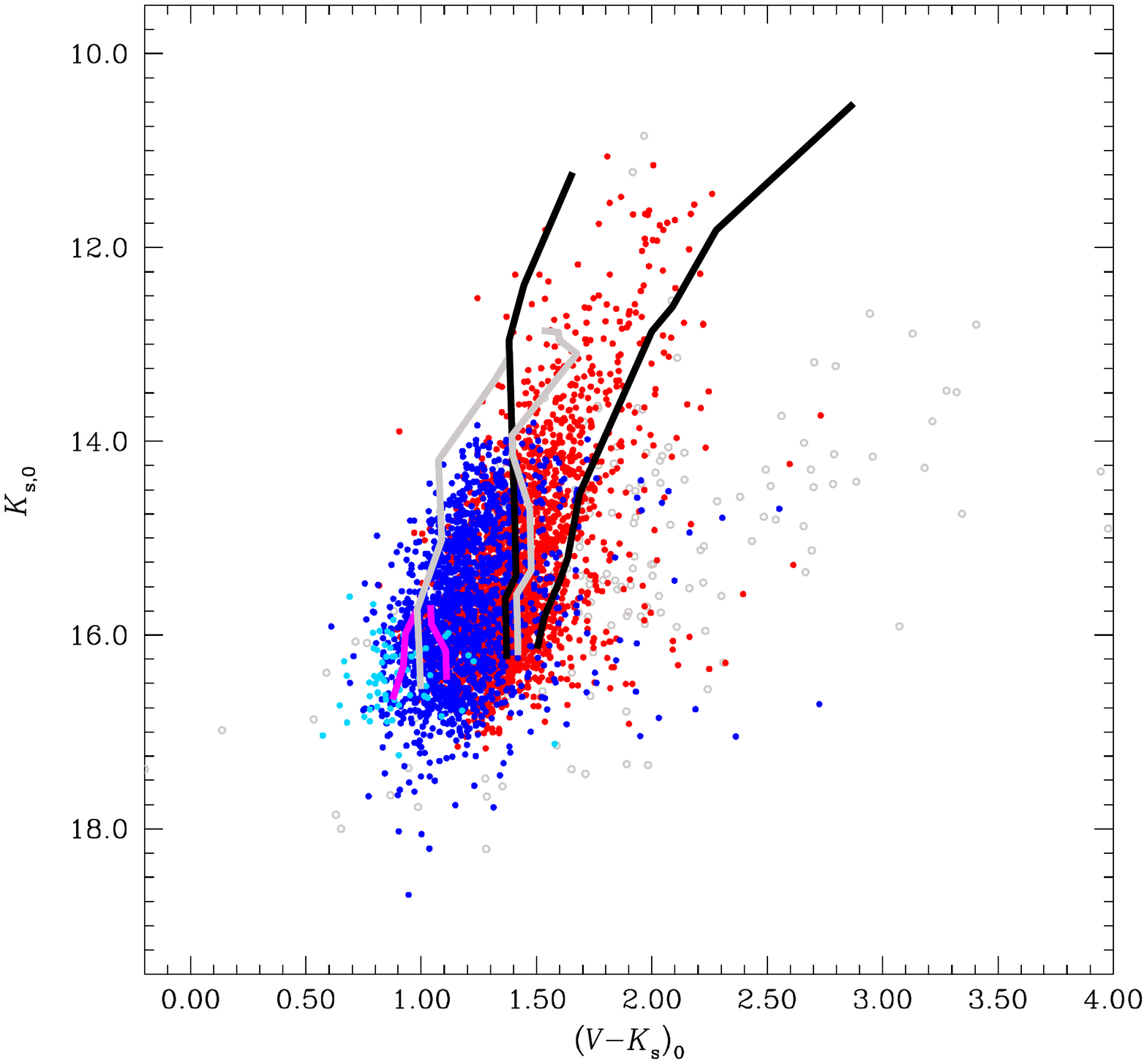}}
\caption{The left, middle, and right panels show the color--magnitude 
  diagrams for the $(K_\mathrm{s}, Y-K_\mathrm{s})$, $(K_\mathrm{s},
  J-K_\mathrm{s})$, and $(K_\mathrm{s}, V-K_\mathrm{s})$ combinations 
  of magnitudes and colors, respectively. The color coding is the same 
  as that in Fig.~\ref{logG}. The middle and right-hand panels also 
  show the theoretical instability strips for F (black lines), FO 
  (gray lines), and SO (magenta lines) CCs, respectively. The models,
  calculated for $Z=0.004$ and $Y=0.25$, have been taken from 
  \citet{Bono2000,Bono2001a,Bono2001b}.\label{hrds}}
 \end{figure*}

 The top and bottom right-hand panels in Figs~\ref{figMC_Y},
\ref{figMC_J}, and \ref{figMC_K} display essentially the same results
as the panels on the left, but for the peak-to-peak amplitudes instead
of the intensity-averaged magnitudes. Again, the results for the
amplitudes in the $K_\mathrm{s}$ band are very good, while the
uncertainties become significantly larger for the $J$ and, especially,
the $Y$ filters.

On the basis of the Monte Carlo experiments, we decided to assign as
uncertainties to the intensity-averaged magnitudes and peak-to-peak
amplitudes the values shown in the bottom panels of
Figs~\ref{figMC_Y}, \ref{figMC_J}, and \ref{figMC_K}.

The light curves and the best-fitting templates found with the
procedure outlined in this section are reported in
Figs~\ref{figY_pag119},~\ref{figJ_pag119}, and \ref{figK_pag119} for
the $Y,\, J$, and $K_\mathrm{s}$ bands, respectively. These figures
display the data for a subsample of 27 CCs; figures including the full
data set of 4172 objects are available in the electronic version of
this paper on the journal's website.

\begin{figure*}
\hbox{
\includegraphics[scale=.4]{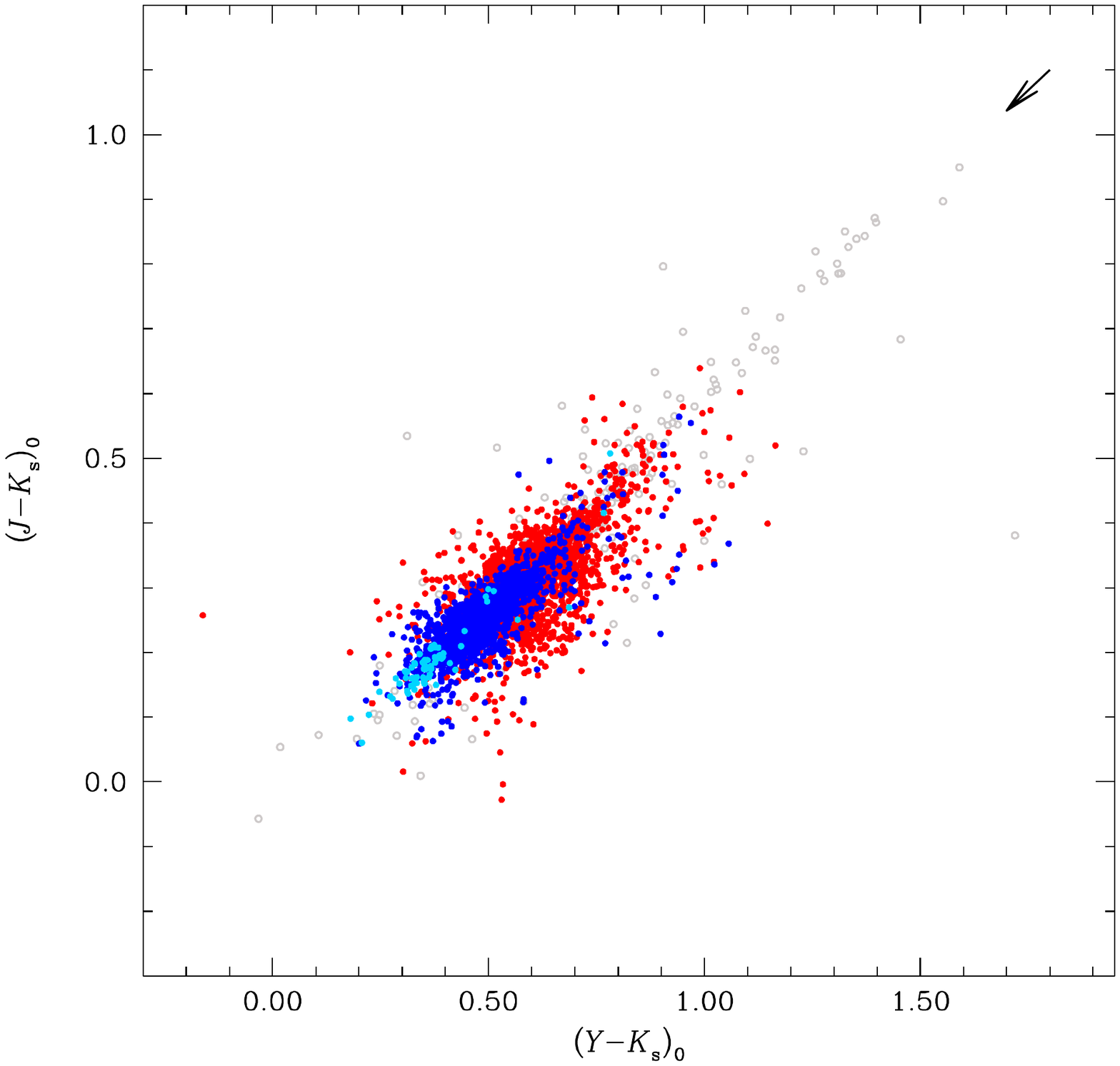}
\includegraphics[scale=.4]{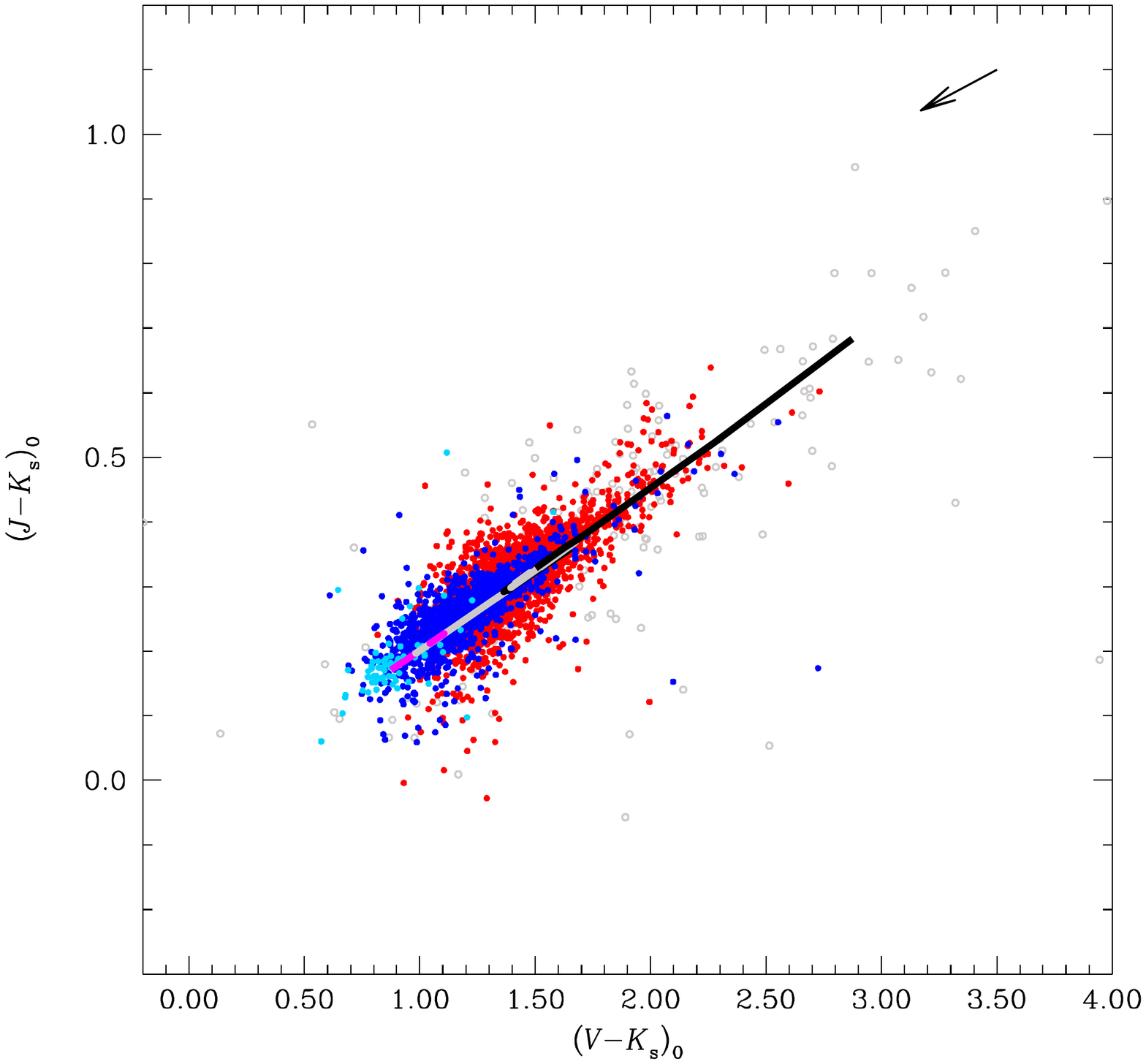}}
\caption{The left- and right-hand panels show the color--color 
  diagrams in the $(J-K_\mathrm{s}, Y-K_\mathrm{s})$ and 
  $(V-K_\mathrm{s}, J-K_\mathrm{s})$ planes, respectively. The color 
  coding is the same as that in Fig.~\ref{logG}. The right-hand panel 
  also shows the theoretical instability strips from 
  Fig.~\ref{hrds}. In both panels, arrows indicate the reddening 
  vectors, calculated for a reddening value $E(V-I)=0.15$ mag, i.e.,
  approximately three times the average reddening in the SMC,
  according to the adopted reddening maps of 
  \citet{Haschke2011}. \label{cc} }
\end{figure*}

 Similarly, Table~\ref{tabResults} reports the main physical quantities
derived on the basis of the fitting procedure, namely the
intensity-averaged magnitude for each variable (and each filter), the
peak-to-peak amplitudes, and the relative errors calculated by means
of the Monte Carlo experiments.

Finally, we recall that the $Y$, $J$, and $K_\mathrm{s}$ photometry
described in this work is defined in the VISTA system. It is possible
to compare our data with measurements in the widely used 2MASS system
\citep[Two Micron All Sky Survey][]{2mass} after applying the system
transformations made available by the Cambridge Astronomy Survey Unit
(CASU)\footnote{http://casu.ast.cam.ac.uk/surveys-projects/vista/technical/photometric-properties}:
($J$$-$$K_\mathrm{s}$)$^{\rm 2M}$=1.081($J$-$K_\mathrm{s}$)$^{\rm V}$,
$J$$^{\rm 2M}$=$J$$^V$+0.07($J$-$K_\mathrm{s}$)$^{\rm V}$, and
$K_\mathrm{s}$$^{\rm 2M}$=$K_\mathrm{s}$$^{\rm
  V}$$-$0.011($J$-$K_\mathrm{s}$)$^{\rm V}$. No transformation is
provided in $Y$ since 2MASS did not observe in this filter. However,
it is possible to ``standardize'' the $Y$ band by applying a color
equation that, at present, is available only as a function of the
$(J-H)$ color (and is therefore of no use to us). A new transformation
using the ($J-K_\mathrm{s}$) bands is being derived by CASU and will
be available within a few months.

\begin{figure}
\epsscale{1.1}
\plotone{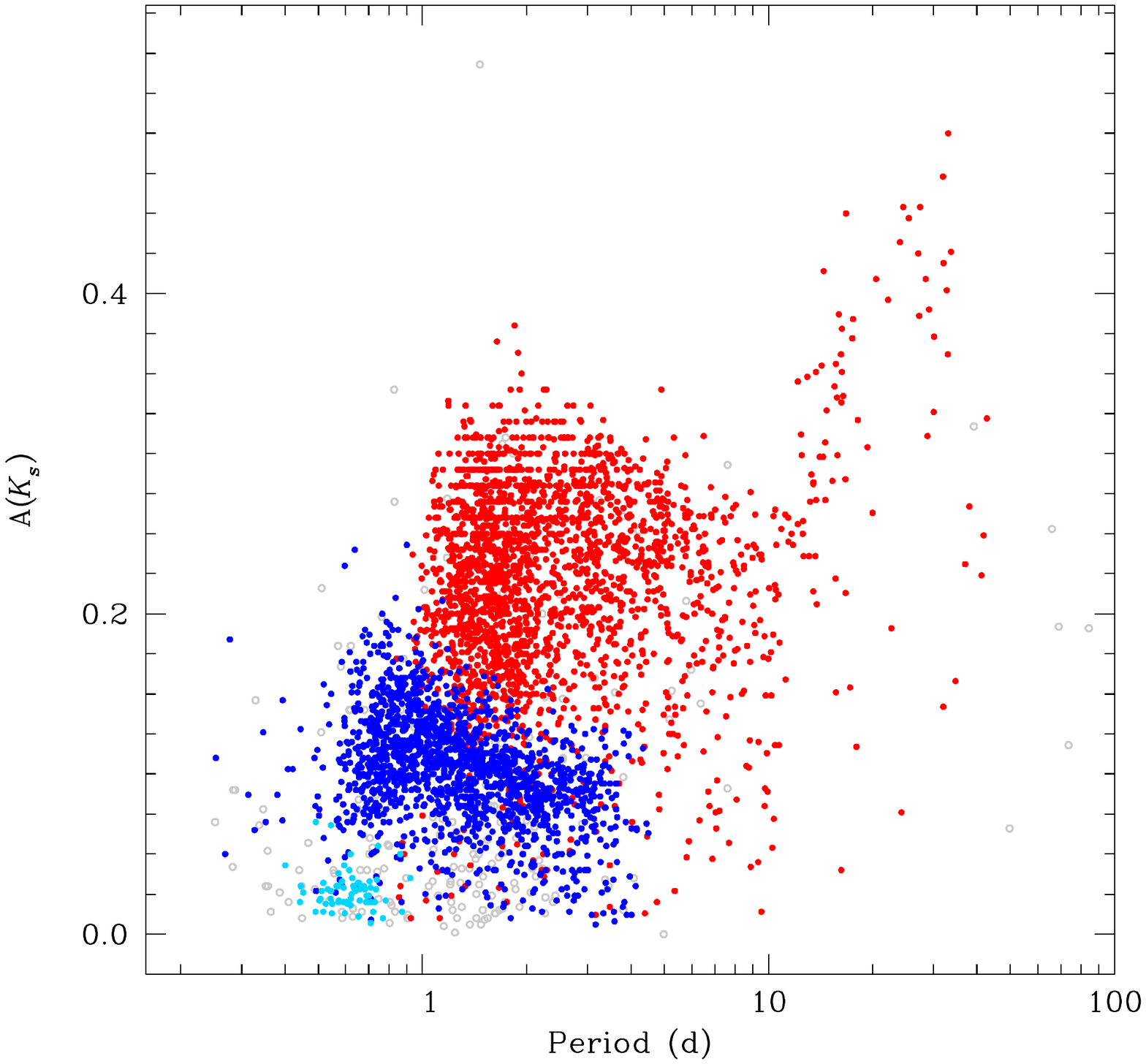}
\caption{Period versus amplitude in the $K_\mathrm{s}$ band for the 
  target CCs. The color coding is the same as that in 
  Fig.~\ref{logG}.  \label{periodAmp}}
\end{figure}

\begin{figure}
\epsscale{1.1}
\plotone{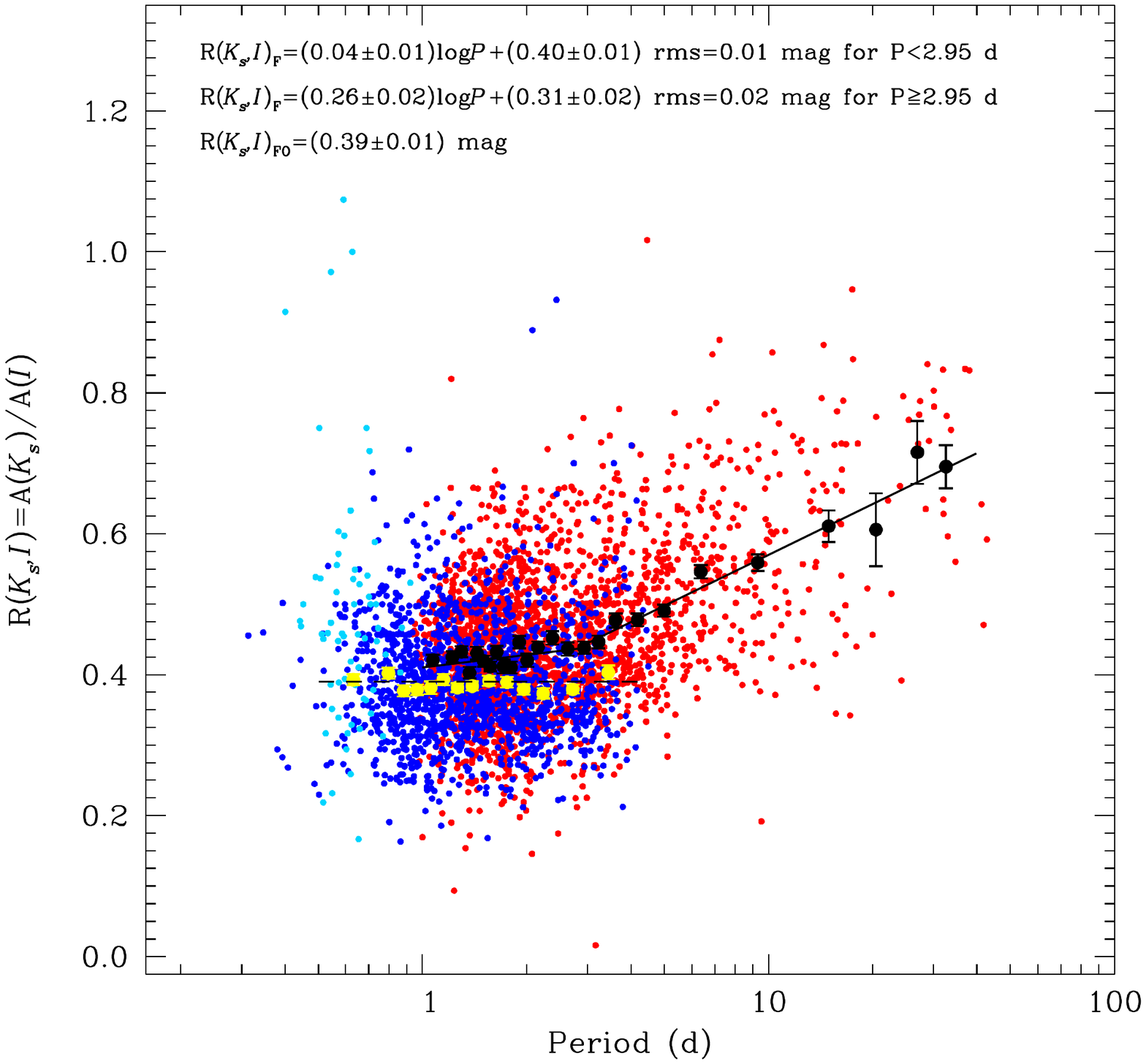}
\caption{Peak-to-peak amplitude ratios A($K_\mathrm{s}$)/A($I$) 
  for the F, FO, and SO mode pulsators studied here. The $I$-band 
  amplitudes band are from the OGLE III survey. The color coding of 
  the small filled circles is as that in Fig.~\ref{logG}. The black 
 and yellow filled circles represent the averages in period bins 
  for F and FO pulsators, respectively. The derived analytical ratios 
  are labelled in the figure. \label{ampRatio}}
\end{figure} 

Since the intrinsic $\langle$$J$ $\rangle$$-$$\langle$ $K_\mathrm{s}$$
\rangle$ colors of the CCs investigated here typically range from 0.1
mag to 0.6 mag, the VISTA and 2MASS $K_\mathrm{s}$ can be considered
equivalent for CCs (see Fig.~\ref{hrds}) to a very good approximation
(better than $\sim$5 mmag), although the corrections needed in the $J$
band can be significant.

\section{Average magnitudes, colors and peak-to-peak amplitudes}

We constructed color--magnitude diagrams for the entire sample of
CC analyzed here, distinguishing them by the different types of
pulsation. The results are shown in Fig.~\ref{hrds}. The middle and
right-hand panels of this figure display the comparison in the
$K_\mathrm{s,0}, (J-K_\mathrm{s})_0$ and
$K_\mathrm{s,0},(V-K_\mathrm{s})_0$ planes of the observed data with
the theoretical instability strips for F, FO, and SO CCs. The models,
calculated for $Z=0.004$ and $Y=0.25$, have been taken from
\citet{Bono2000,Bono2001a,Bono2001b}. We note that the models are in
the $JHK$ Johnson system. Thus, we have to converted them into the
VISTA system, adopting the VISTA--2MASS relations referred to in the
previous section, as well as the color transformations available from
\citet{Bessell1988} and \citet{Carpenter2001}. As a result, we
obtained the following approximate equations:
\begin{eqnarray}
K_\mathrm{s}^{\rm V} & = & K^{\rm J}+0.007\,(V-K)^{\rm
  J}+0.03\,(J-K)^{\rm J}+ \nonumber\\
 &  & -0.038 \\
 (V-K_\mathrm{s})^{\rm V} & = & 0.993\,(V-K)^{\rm J}-0.03\,(J-K)^{\rm
   J}+ \nonumber \\
 &  & +0.038 \\
 (J-K_\mathrm{s})^{\rm V} & = & 0.87\,(J-K)^{\rm J}-0.01 
\end{eqnarray} 
\noindent 
where the superscripts ``V'' and ``J'' refer to quantities in the
VISTA and Johnson systems, respectively. There is general good
agreement between predicted colors and observations, especially for FO
pulsators, while for F pulsators the observed instability strip
appears to be larger at low luminosities (i.e., short periods)
compared with the predictions.

\begin{figure*}
\epsscale{0.85}
\plotone{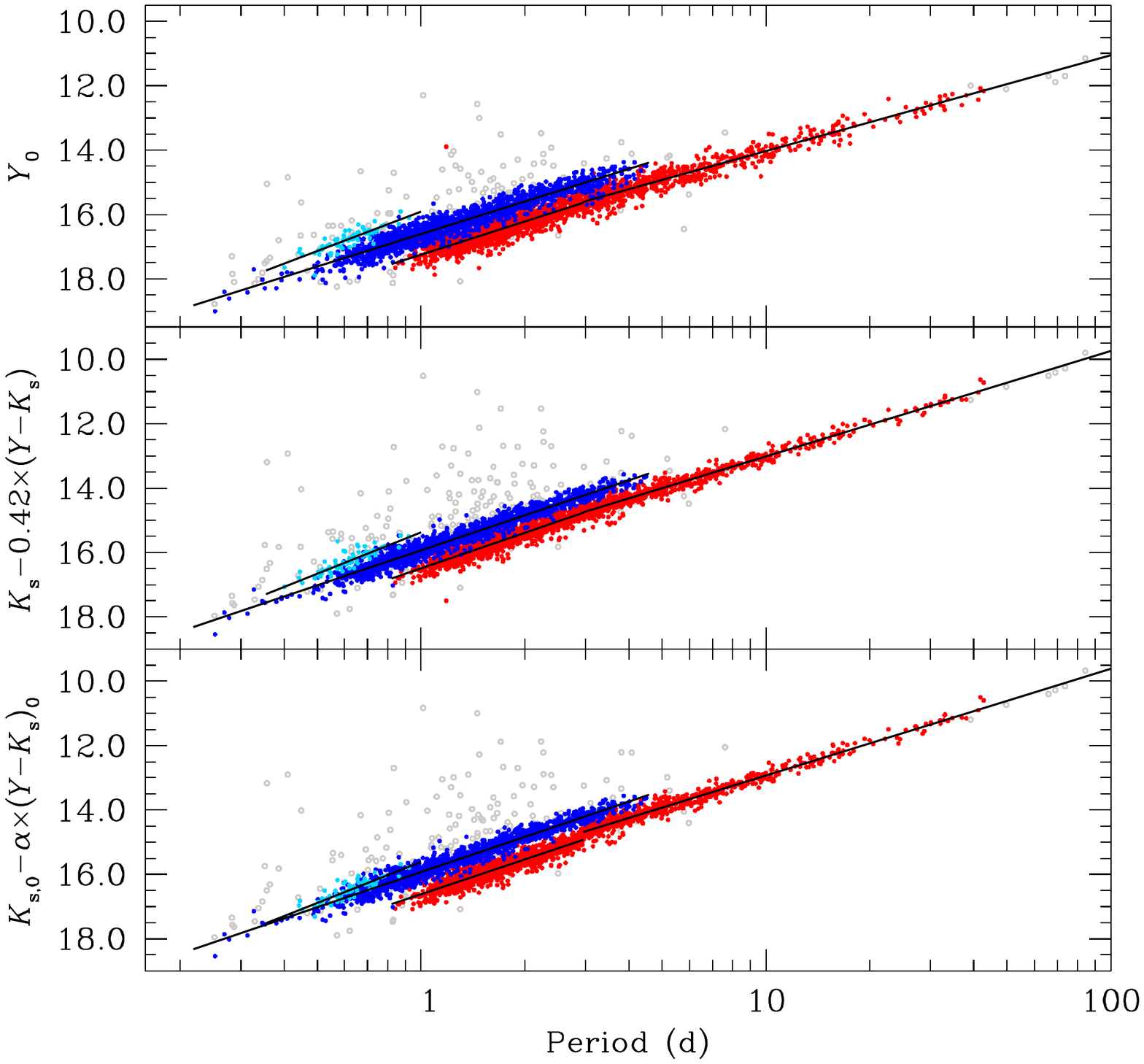}
\caption{From top to bottom, $PL(Y)$, $PW(K_\mathrm{s},Y)$, and 
  $PLC(K_\mathrm{s},Y)$ relations for the SMC CCs investigated in this 
  paper. The color code is the same as that in Fig.~\ref{logG}. The 
  solid lines represent least-squares fits to the data shown in 
  Table~\ref{pl}. Note that the discontinuity in the bottom panel,
  both in data and fit, for F-mode pulsators is only due to the 
  visualization procedure (projection from 3 to 2 
  dimensions).\label{ply}}
\end{figure*}

\begin{figure*}
\epsscale{0.85}
\plotone{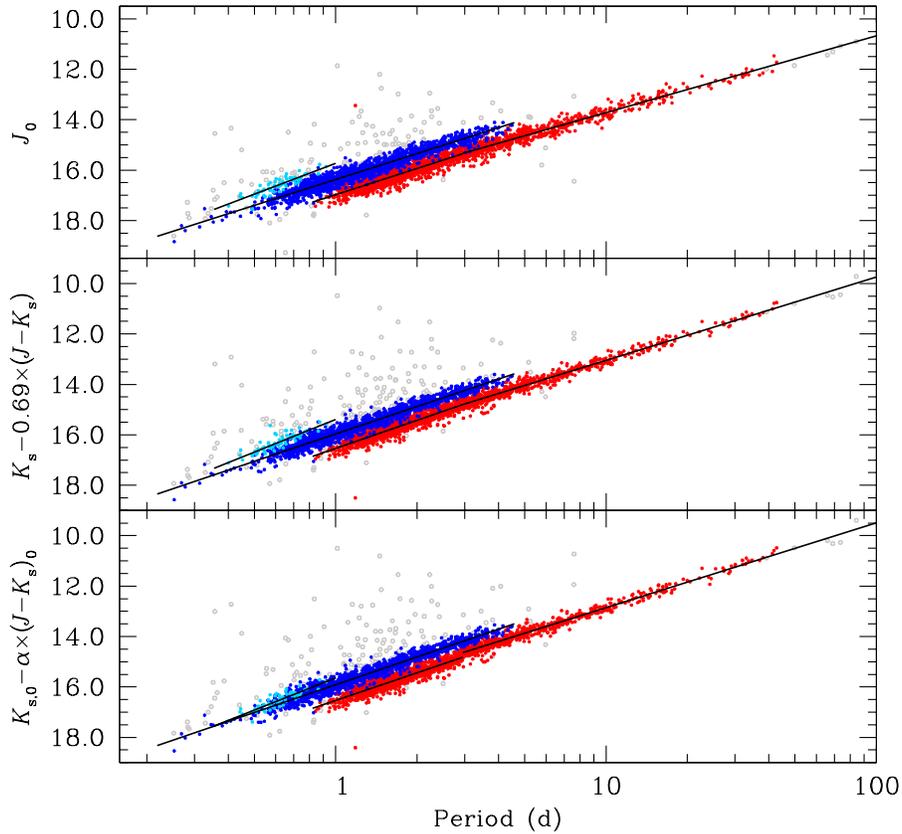}
\caption{As Fig.~\ref{ply} but for $PL(J)$, $PW(K_\mathrm{s},J)$, and 
  $PLC(K_\mathrm{s},J)$.\label{plj}}
\end{figure*}

\begin{figure*}
\epsscale{0.85}
\plotone{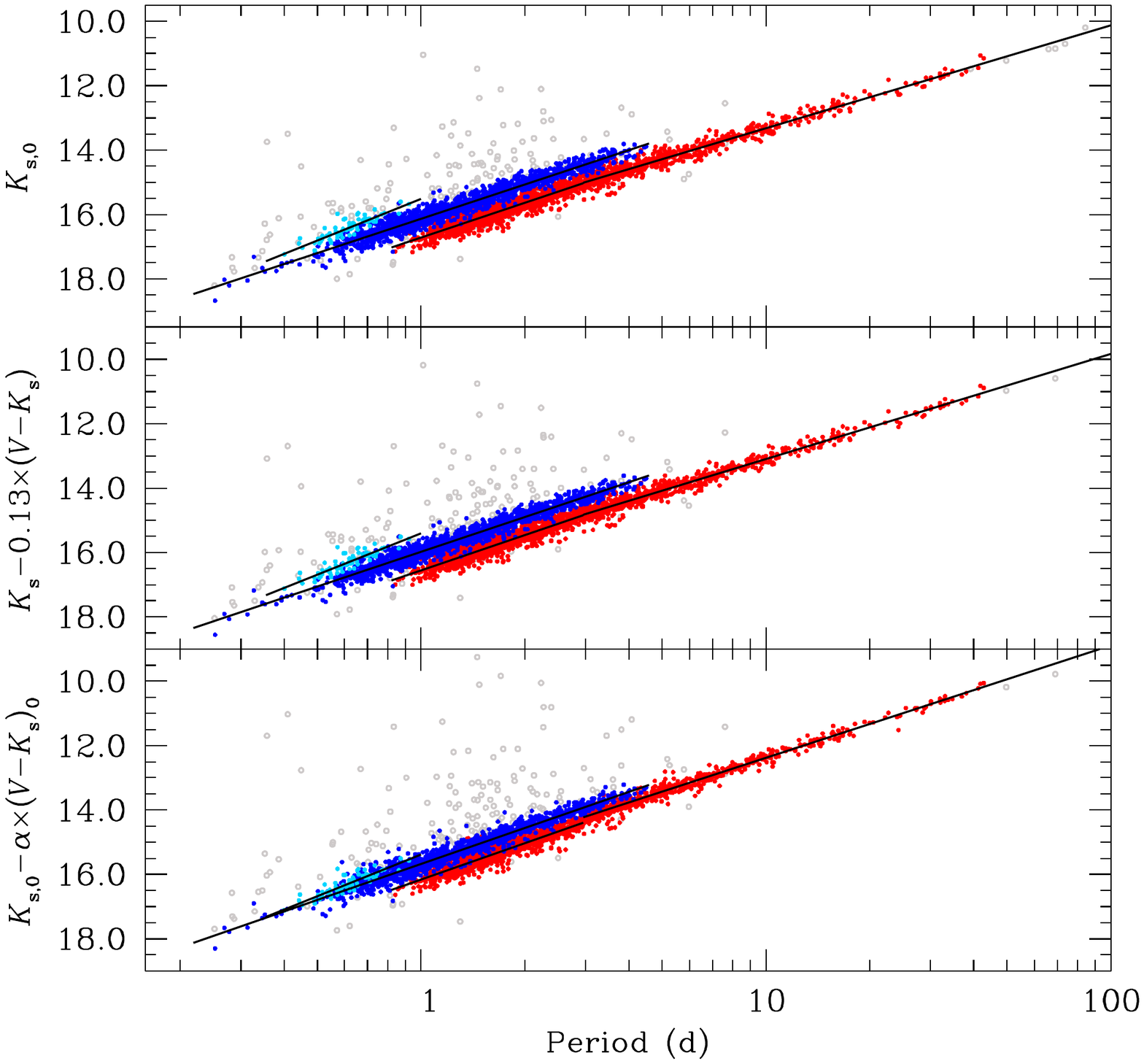}
\caption{As Fig.~\ref{ply} but for $PL(K_\mathrm{s})$,
  $PW(K_\mathrm{s},V)$, and $PLC(K_\mathrm{s},V)$. \label{plk}}
\end{figure*}

Additional information can be obtained from the color--color diagrams
shown in Fig.~\ref{cc}, where the left- and right-hand panels show the
$(J-K_\mathrm{s}, Y-K_\mathrm{s})$ and $(V-K_\mathrm{s},
J-K_\mathrm{s})$ planes, respectively. In both cases, arrows represent
the reddening vectors, which are almost parallel to the data
distribution, making it almost impossible to use these planes to
estimate individual reddening values. It is interesting to note the
distribution of the rejected stars (empty gray circles), which is
markedly elongated towards red colors (especially in the
$(J-K_\mathrm{s}, Y-K_\mathrm{s})$ plane). This suggests significant
contamination in the $K_\mathrm{s}$ band by very red objects, likely
red clump or red/asymptotic giant-branch stars, or bright (early-type)
background galaxies. The right-hand panel of Fig.~\ref{cc} shows the
theoretical instability strip, now visible as an almost straight line
passing, as expected, through the middle of the data
distribution. Indeed, at fixed effective temperature, the position in
the color--color plane is unequivocally determined by the adopted
effective temperature--color transformation. Note that the modest
broadening of the data ($\sim$0.07-0.1 mag) is due to different contributions, namely the
photometric errors, blending effects, and/or small metallicity
differences, but not to reddening effects (see the direction of the
reddening vectors in both planes of Fig.~\ref{cc}).

Figure~\ref{periodAmp} shows the period versus peak-to-peak amplitude
in the $K_\mathrm{s}$ band for the target CCs. As far as we know, this
is the first time that such a plane has been exploited with such a
statistically significant number of objects in an infrared band. An
inspection of the figure reveals the clean separation in amplitude of
the three different modes. It is interesting to note that the peculiar
shape of the distribution of F pulsators, with an increase at about
$P=10$ d and a maximum around $P=22$--24 d, resembles a similar trend
observed in the visual $V$ band for the Galactic CCs \citep[][]{Bono2000}
and is consistent with theoretical predictions (see their Fig. 7).

We also looked at the peak-to-peak amplitude ratios for different
pulsation modes between the $K_\mathrm{s}$ and $I$ bands. These values
may be useful for authors who want to use the canonical
template-fitting procedure. Figure~\ref{ampRatio} shows the
$R$($K_\mathrm{s},I$)= A($K_\mathrm{s}$)/A($I$) ratio versus
period for the CCs investigated here. We calculated the ratio between
these bands, because our amplitudes are more accurate in
$K_\mathrm{s}$ relative to $Y$ and $J$ (see Section~\ref{template})
and OGLE\,III provides the peak-to-peak amplitudes for all Cepheids
investigated here only in the $I$ filter.

Given the rather large scatter in the data (possibly in part due to
the presence of binary companions), we decided to average the CCs in
period bins, obtaining the light blue and yellow filled circles for F
and FO pulsators, respectively (we did not consider the SO CCs because
of their very small amplitudes). An analysis of the averaged data
reveals the different behavior of F and FO
pulsators. $R$($K_\mathrm{s},I$) is almost constant for FO pulsators
over the full period range, while for F pulsators it is flat only
until $P\sim2.95$ d. For longer periods there is a steep linear
increase of $R$($K_\mathrm{s},I$) with period. Quantitatively, we
derived the following equations for F and FO pulsators:

\begin{eqnarray} 
 R(K_S,I)_F & = & (0.04\pm0.01)\log P+(0.40\pm0.01) \nonumber \\
  &  & (P<2.95~{\rm d}) \label{eqnRatios1} \\
R(K_S,I)_F & = & (0.26\pm0.02)\log P+(0.31\pm0.02)  \nonumber \\
 &  & (P\ge2.95~{\rm d})  \label{eqnRatios2} \\
 R(K_S,I)_{FO} & = &(0.39\pm0.01).  \label{eqnRatios3}
\end{eqnarray}

\noindent 
It is interesting to note that the steep change in slope for F
pulsators occurs at about the same period where we find a break in the
$PL$, $PW$, and $PLC$ relations (see next section).

A comparison of our results with those in the literature reveals some
differences. Indeed, concerning F pulsators only,
\citet{Soszynski2005} suggest using constant values of $R(K_S,I)$=0.49
or 0.62 for periods $<$/$\geq$\,$\sim$20 d, respectively. Using
Eq.~\ref{eqnRatios2}, for $P \sim 20$ d we obtain $R(K_S,I)$=0.65,
which is fully compatible with \citet{Soszynski2005}'s
values. However, it is easy to verify that the agreement is worse for
different periods. For example, at $P=40$ d, we obtain
$R(K_S,I)$=0.73, while at $P=2$ d $R(K_S,I)$=0.41. Taking into
  account that the \citet{Soszynski2005} results have been derived using
 Galactic and LMC CCs, it is reasonable to hypothesize that part of
 the discrepancy between our and their findings  can be owing to
the different metallicities of the adopted CC samples.

We cannot perform a direct comparison with \citet{Inno2015}'s results,
because they only provide the ratio of NIR bands with respect to the
$V$ band. However, we can compare the trends versus the periods, since
they have different data sets for Galactic+LMC and SMC CCs. As a
result, \citet{Inno2015} found a break in $R(K_S,I)$ at a period
similar to that of \citet{Soszynski2005}. This is in contrast with our
results (perhaps this is due to the smaller size of their sample).  On
the other hand, they found systematically lower $R(K_S,I)$ values for
SMC CCs, with respect to the Galactic+LMC variables, in full agreement
with our results.

\section{$PL$, $PLC$, and $PW$ relations}
\label{relations}

The data reported in Table~\ref{tabResults} allow us to derive several
useful relationships, adopting various combinations of magnitudes and
colors. In particular, we derived $PL$ relations in $Y$, $J$, and
$K_\mathrm{s}$, as well as $PW$ and $PLC$ relations for the following
combinations: ($K_\mathrm{s}$,$Y-K_\mathrm{s}$),
($K_\mathrm{s}$,$J-K_\mathrm{s}$), and
($K_\mathrm{s}$,$V-K_\mathrm{s}$).

Before deriving the latter relationships we have to take account of
the reddening. We adopted the extinction maps of \citet{Haschke2011},
as we already successfully did in our previous papers \citep[see,
  e.g., the discussion in Sect. 3 of][]{Ripepi2015}.
The reddening values were converted using the following equations:
$E$($Y$$-$$K_\mathrm{s}$)=1.80$E$($V$$-$$I$);
$E$($V$$-$$K_\mathrm{s}$)=2.24$E$($V$$-$$I$); 
$E$($J$$-$$ K_\mathrm{s}$)=0.43$E$($V$$-$$I$)
\citep{Cardelli1989,Kerber2009,Gao2013}.\footnote{The coefficients
  used in this paper are consistent with the 2MASS system, to which
  the VISTA system is related.} The coefficients of the $PW$
relations were calculated in a similar fashion.

To derive the relevant relationships for F, FO, and SO variables, we
adopted equations of the form $M1_0=\alpha+\beta \log P$,
$W(M1,M2)=\alpha+\beta \log P$, and $M1_0=\alpha+\beta \log P+\gamma
(M2-M1)_0$ for the $PL$, $PW$, and $PLC$ relations,
respectively. Here, $M1$ and $M2$ represent two different magnitudes
from among $V$, $Y$, $J$, and $K_\mathrm{s}$. The details about the
combinations of magnitudes and colors adopted in this papers can be
found in Table~\ref{pl}. In order to use the full sample of pulsators,
including the double- or multi-mode CCs, we decided to use them with
the period of the dominant mode (e.g., F-mode if the star is an F/FO
double-mode pulsator, and so on). This procedure is safe, since from
our previous investigation of LMC pulsators \citep{Ripepi2012a} we
know that these objects do not exhibit systematic luminosity
differences with respect to single-mode objects.

The next step involved checking for the presence of changes in the
slope of the different relationships, as found in previous studies in
the literature \citep[see, e.g.][and references
  therein]{Subramanian2015}. To this aim, we used the $PW$ in
$V,K_\mathrm{s}$ which was known from our previous investigation of
the LMC CCs, to show a small intrinsic dispersion \citep[see,
  e.g.][]{Ripepi2012a}, and which is thus particularly appropriate for
our purpose. As a result, we found that there is a clear change in
slope at $\log P=0.47$ ($\sim$ 2.95 d) for F-mode pulsators, while
there is no significant change in slope for FO variables. This was
confirmed by the analysis of the $PL$ and $PW$ in different filters
and is in agreement with the results obtained in the optical ($V, I$
bands) for the OGLE\,III sample of F-mode pulsators by
\citet[][]{Subramanian2015} (see also references in this
paper). However, we do not find the break at $\log P=0.029$ ($\sim$
1.07 d) that they noticed for FO-mode pulsators. A possible
explanation for the break detected at $P\sim$3 d is that for shorter
periods the blue loop of the Cepheid evolutionary track is too short
and enters only the reddest part of the instability strip (M. Marconi
et al., in prep).

\begin{figure}
\epsscale{1.2}
\plotone{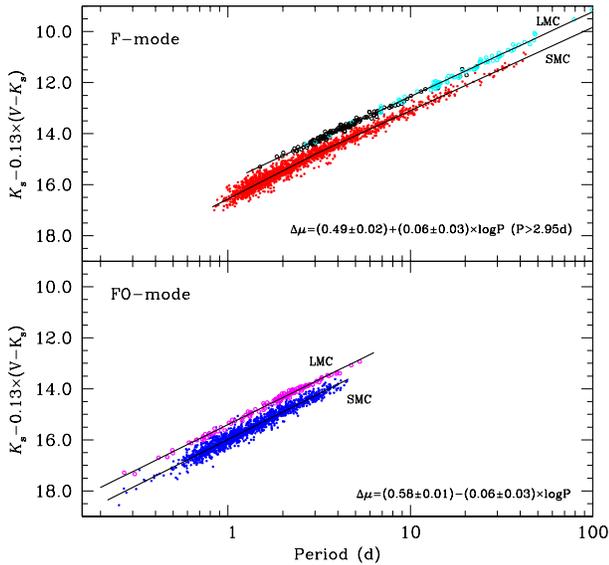}
\caption{Top: $PW(K_\mathrm{s},V)$ for the LMC and SMC F-mode CCs 
  according to \citet{Ripepi2012a} (black circles),
  \citet{Persson2004} (light blue circles), and this paper (filled red 
  circles), respectively. Bottom: as the top panel but for FO-mode 
  pulsators in the LMC \citep[magenta circles,][]{Ripepi2012a} and in 
  the SMC (filled blue circles, this work). In both panels the solid 
  lines represent the best fits to the data (see text for details). 
\label{figMC}}
\end{figure}

The $PW(V,K_\mathrm{s})$ was also used to analyze problematic objects,
identified as clear outliers from these relations. In total, we
discarded 223 CCs. We identified different (but often concurrent)
reasons for the erratic behavior of these objects (see the final
column in Table~\ref{tabResults} for details): (i) misidentification:
all objects with separation VMC--OGLE III $>$ 0.2$''$ were visually
inspected and rejected if they were found to be overluminous in the
$PW(V,K_\mathrm{s})$ relation (more than 100 objects were rejected as
such); (ii) scattered or heavily undersampled light curves (always low
$G$/high $\chi^2$ values; more than 50 such objects were present);
(iii) notes from either OGLE\,III or VMC, i.e., the presence of flags
reporting problems with the images (more than 20 rejections); (iv)
saturation (seven objects). Note also that 17 objects with good VMC
photometry were rejected because they lacked OGLE\,III $V$-band
photometry. Not all outliers can be explained by invoking these
reasons; in fact, there are 35 outliers for which we could find no
apparent flaws. However, most are faint and all are overluminous.
Hence, it is likely that they suffer from blending with bright
neighbor stars. The discarded stars are reported separately, both in
Table~\ref{tabResults} and Figs~\ref{figY_pag119},~\ref{figJ_pag119},
and~\ref{figK_pag119}. Finally, we note that a few other objects were
excluded from the derivation of the $PL$, $PW$, and $PLC$ relations
involving the $Y$ or $J$ bands because of specific problems in these
bands. To avoid confusion, these objects have not been highlighted in
the table and figures.

On this basis, we performed a least-squares fit to the data to derive
all relations, adopting a break at $\log P=0.47$ ($P\sim$2.95 d) for
F-mode CCs, while all FO- and SO-mode pulsators were used together. The
results of this work are shown in Table~\ref{pl} and
Figs~\ref{ply},~\ref{plj}, and~\ref{plk}, where from top to bottom we
display the F-, FO-, and SO-mode $PL$, $PW$, and $PLC$ relations,
respectively. Note that the $PLC$ relations show a distinct
discontinuity at $\log P=0.47$ owing to the way the data are projected
in two dimensions. As far as we know, these are the first CC $PL$,
$PW$, and $PLC$ relations ever derived that involve the $Y$ band. The
same is true for SO pulsators, even if in this case the small number
of objects available for the calculations (about 70) and their
intrinsic faintness did not allow us to obtain $PL$, $PW$, and $PLC$
relations to similar precision as those for F and FO pulsators (see
Table~\ref{pl}).

The regressions listed in Table~\ref{pl} (and in
Figs~\ref{ply},~\ref{plj}, and~\ref{plk}) show that the $PL$ relations
have, as expected, a larger dispersion with respect to the $PW$ and
$PLC$ relations, which show a similar scatter for all combinations of
magnitudes and colors, even though the use of $(V-K_\mathrm{s})$ and
$(J-K_\mathrm{s})$ give slightly better results. This is not
surprising, since the general quality of the $Y$-band data is
(moderately) worse than that in $J$. In any case, the constancy of the
dispersion of all these relations is a clear indication that the
elongated structure of the SMC is dominating the intrinsic dispersion
of these relations, which we know from LMC studies to be much smaller
\citep[see, e.g.][]{Ripepi2012a,Inno2013,Macri2015}.

We can now compare our results with the previous investigation by
\citet{Inno2013}. These authors derived $PW$ relations for a variety
of combinations of magnitudes and colors for SMC and LMC CCs,
including the NIR bands $J,H, K_\mathrm{s}$. Their photometric
database relies mainly on single-epoch light curves, from which they
derived average magnitudes by adopting some literature template light
curves and relying on published ephemerides and amplitude ratios
(e.g., A($J$)/A($I$)). It is important to note our very different
approach with respect to theirs. Indeed, the larger number of observed
epochs (especially in the $K_\mathrm{s}$ band) allowed us to adopt a
template procedure without having to rely on any external information
(apart from the periods, see details in Section~\ref{template}) and
which is capable of achieving much higher precision of the
intensity-averaged photometry for each individual CC. The relations we
can compare with \citet{Inno2013} are the $PW(J,K_\mathrm{s})$ and
$PW(V,K_\mathrm{s})$ for F and FO pulsators. The latter authors
calculated these relations in different ways, either without taking
into account any break or by arbitrarily imposing breaks at $\log
P=0.35,0.40,0.45$. Therefore, we can compare the $PW(J,K_\mathrm{s})$
and $PW(V,K_\mathrm{s})$ relations with no break for FO pulsators
\citep[Table 1 of][]{Inno2013} and the $PW(J,K_\mathrm{s})$ relation
for F pulsators with a break at $\log P \geq 0.45$ \citep[see Table 3
  of][]{Inno2013}\footnote{Note that \citet{Inno2013}'s
  $PW(V,K_\mathrm{s})$ relations are not provided for different
  breaks, nor do they have relations for $\log P > 0.45$}. To take
into account that our photometry is in the VISTA system, while
\citet{Inno2013}'s were in the 2MASS system, we have applied the
equations discussed in Section~\ref{template} to convert
\citet{Inno2013}'s relations to the VISTA system. We can now finally
perform the comparison with the values listed in our Table~\ref{pl}.
We obtain very good agreement for the three $PW$ relations quoted
above, in all cases within $\sim\, 1\, \sigma$. However, we emphasize
that the precision for the individual CC $W$ magnitudes is better in
our case given the larger number of observations. This is an important
factor when dealing with the structure of the SMC, whose study
requires precise individual relative distances.

\subsection{The relative distance between SMC and LMC and the absolute
distance of the SMC}

The relationships derived in the previous Section will be used in a
forthcoming paper to study in detail the 3D structure of the
SMC. However, a first important use of the data presented in this
paper is the estimation of the relative distance between the two
MCs. In turn, the assumption of a distance for the LMC, which can be
more safely determined with respect to the SMC's (since the SMC is so
significantly elongated), allows us to provide an estimate of the
absolute distance to the SMC (or, rather, of the center defined by the
CC distribution).

We hence proceeded using our own data published in \citet{Ripepi2012a}
for the CCs in the LMC. This is justified because (i) we used data in
the same photometric system, (ii) we obtained a $PW(V,K_\mathrm{s})$
relation with very low dispersion for the LMC CCs, and (iii) we also
provided an absolute distance estimate for the LMC.

The technique adopted is illustrated in Fig.~\ref{figMC}, where we
compare the $PW(V,K_\mathrm{s})$ relations for F- and FO-mode
pulsators (top and bottom panels, respectively). First, observing the
period distribution of the CCs in the LMC and the fact that the slope
of the LMC's $PW(V,K_\mathrm{s})$ relation is very close to the slope
we have found here for the SMC CCs characterized by $\log P>0.47$, we
used this latter relation for our comparison of F-mode
pulsators. Nonetheless, the slopes of the relations for both F- and
FO-mode pulsators are slightly different for the LMC and SMC (which is
possibly related to a weak but significant metallicity
dependence). Indeed, it is possible to describe the difference in
$W(V,K_\mathrm{s})$, which translates directly into a difference in
distance modulus $\mu$ as a function of period with two simple
equations:

\begin{eqnarray}
\Delta\mu_{\rm F} &= & (0.49\pm0.02)+(0.06\pm0.03)\,\log P \label{deltamu1}\\
\Delta\mu_{\rm FO} &= & (0.58\pm0.01)-(0.06\pm0.03)\,\log P  \label{deltamu2}
\end{eqnarray}
\noindent
where $\Delta\mu$ means the difference in distance modulus of SMC and
LMC, and the errors take into account the uncertainties in both the
LMC and SMC relations. To use Eq.~\ref{deltamu1} and ~\ref{deltamu2},
we have to fix pivoting periods to determine the $\Delta\mu$
values. After some tests we chose $P=10$ d and $P=2$ d for F- and
FO-mode pulsators, respectively. These values are approximately in the
middle of the period range for both pulsator types, but it is easy to
verify that the results do not depend significantly on this
choice. The result of this exercise gives: $\Delta\mu_{\rm
  F}=0.55\pm0.04$ mag and $\Delta\mu_{\rm FO}=0.56\pm0.03$ mag, in
excellent mutual agreement. Averaging the two results we obtain our
best estimate for the relative distance between the MCs:
$\Delta\mu=0.55\pm0.04$ mag. This value is in good agreement with that
derived in a similar fashion by \citet{Inno2013}, especially with
their result for FO pulsators: $\Delta\mu=0.52\pm0.03$ mag, while for
F-mode CCs they find $\Delta\mu=0.48\pm0.03$ mag. Our estimate is
somewhat larger than those quotes in other papers based on different
standard candles \citep[see][for a large compilation of distance
  differentials]{degrijs2014}. For example, \citet{Cioni2000b} found
$\Delta\mu=0.44\pm0.05$ from the tip of the red-giant branch, while
using RR Lyrae stars \citet{Szewczyk2009} found a significantly
smaller value, $\Delta\mu= 0.327\pm0.002$ mag. According to
\citet{Matsunaga2011}, Type II Cepheids (W Vir) yield
$\Delta\mu=0.40\pm0.07$ or $\Delta\mu=0.39\pm0.05$ mag (depending on
the use of NIR or optical data, respectively). In general, Table 4 of
\citet{Matsunaga2011}, where they list several literature results,
seems to suggest that all evaluations of the $\Delta\mu$ based on CCs
provide larger values with respect to those based on population II
indicators. This can be due to the very different spatial distribution
among CCs (typically showing a disk-like location in both MCs) and
population II tracers (e.g., RR Lyrae stars, type II Cepheids, which
are more evenly distributed around a sort of spheroid in both MCs), as
shown, e.g., by \citet{Deb2014,Moretti2014,Deb2015}.

The absolute distance to the SMC can be determined by simply adding to
the $\Delta\mu$ estimated above the preferred absolute distance for
the LMC. There are hundreds of such estimates in the literature
\citep[see][for a thorough review]{degrijs2014}, but here we will
consider in particular two values: (1) $\mu_{\rm LMC}=18.46\pm0.03$
mag obtained in our previous work on LMC CCs \citep[][]{Ripepi2012a},
and (2) $\mu_{\rm LMC}=18.49\pm0.05$, accurately estimated by
\citet{Pietrzynski2013} on the basis of an eclipsing Cepheid binary
star. As a result, we obtain: (1) $\mu_{\rm SMC}=19.01\pm0.05$ mag and
(2) $\mu_{\rm SMC}=19.04\pm0.06$ mag. These values are formally in
agreement within $\sim\,1\sigma$ with that obtained in
\citet{degrjis2015} by averaging a large number of literature
estimates: $\mu_{\rm SMC}=18.96\pm0.02$ mag. However, as noted by
these latter authors, the systematic uncertainty on this
determination, caused by different sources (mainly the complex SMC
geometry and its elongation along the line of sight), can be as large
as 0.15--0.20 mag.

\section{Conclusions}

In this paper we have presented the VMC survey's light curves for 4172
CCs in the SMC. The majority of the objects have optical $V,I$ data as
well as identification and periods from the OGLE\,III survey, while 13
CCs have been identified by the EROS\,2 survey. Our data set consists
of $Y$, $J$, and $K_\mathrm{s}$ light curves with the number of epochs
typically ranging from 4 to 12 in $Y$ and $J$, and 13 to 36 in
$K_\mathrm{s}$. We used our best light curves in each filter to
construct samples of eight templates covering the full variety of
periods and light-curve shapes. These templates have been used to
automatically perform least-squares fits to the observations by
varying both the amplitude and the phasing, and eventually choosing
the best-fitting template by means of appropriately chosen parameters.
We provide intensity-averaged magnitudes and peak-to-peak amplitudes
in the $Y$, $J$, and $K_\mathrm{s}$ filters. To estimate reliable
uncertainties on these values, we carried out Monte Carlo simulations,
producing 100 mock light curves, adding Gaussian errors to the actual
data for each CC in each filter, and running the template-fitting
procedure from scratch each time. This process allowed us to assess
the reliability of our template-fitting procedure and estimate robust
uncertainties on CC magnitudes and amplitudes.

The intensity-averaged magnitudes in the VISTA $Y$, $J$, and
$K_\mathrm{s}$ filters have been complemented with optical $V$-band
data and periods to construct a variety of $PL$, $PW$, and $PLC$
relations for the CCs in the SMC. The relations involving $V$, $J$,
and $K_\mathrm{s}$ are in agreement with those in the literature. As
for the $Y$ band, to our knowledge in this paper we present the first
CC $PL$, $PW$, and $PLC$ ever obtained using this filter. The $PL$,
$PW$, and $PLC$ relations in the $V, J$, and $K_\mathrm{s}$ bands for
F- and FO-mode CCs in the SMC presented here are the most accurate to
date, since they are based on well- or moderately well-sampled light
curves in $K_\mathrm{s}$ and $J$, respectively. We also presented the
first NIR $PL$, $PW$, and $PLC$ relations for SO pulsators to date.

We used the $PW(V,K_\mathrm{s})$ relation to estimate the relative
SMC--LMC distance and, in turn, the absolute distance of the SMC. For
the former, we derive $\Delta\mu=0.55\pm0.04$ mag, a value that is in
rather good agreement with other evaluations based on CCs, but in
disagreement (significantly larger) with estimates based on (old)
population II distance indicators. We speculate that this discrepancy
may be mainly due to the different geometric distribution of young and
old tracers in the MCs. As for the absolute distance to the SMC, our
best estimates, $\mu_{\rm SMC}=19.01\pm0.05$ mag and $\mu_{\rm
  SMC}=19.04\pm0.06$ mag, based on two particular evaluations of the
distance to the LMC, are in good agreement with literature
values. However, we have to take into account the large systematic
uncertainty due to the complex geometry of the SMC. In a forthcoming
paper, we will use our precise $PW$ relations to unveil the 3D
structure of the SMC. For the reasons outlined above, this work is
also expected to reduce the systematic uncertainties associated with
the absolute distance to the SMC.

\acknowledgments 

We thank our anonymous referee for his/her pertinent and helpful comments.  
This paper is based on observations taken with the ESO/VISTA telescope
located at Paranal (Chile). V. R. warmly thanks Roberto Molinaro for
providing the software for the spline interpolation and the Fourier
analysis used to construct the template light curves. Partial
financial support for this work was provided by PRIN MIUR 2011 (PI: F.
Matteucci). We thank the UK’s VISTA Data Flow System comprising the
VISTA pipeline at CASU and the VISTA Science Archive at Wide Field
Astronomy Unit (Edinburgh; WFAU) for providing calibrated data
products supported by the STFC. This work was partially supported by
the Gaia Research for European Astronomy Training (GREAT-ITN) Marie
Curie network, funded through the European Union Seventh Framework
Programme ([FP7/2007-1312 2013] under grant agreement
no. 264895). M.-R. L. C. acknowledges support from the German Exchange
service. This work is supported by STFC grants ST/5001333 and
ST/M001008. R. d. G. acknowledges financial support from the National
Natural Research Foundation of China (grant 11373010).




\vspace{10cm}

\clearpage 



\clearpage

\begin{deluxetable}{cccc}
\tabletypesize{\scriptsize}
\tablecaption{Number of CCs in each VMC SMC tile.\label{ncep1}}
\tablewidth{0pt}
\tablehead{
\colhead{Tile} & \colhead{RA} & \colhead{DEC} & \colhead{N} \\
\colhead{} & \colhead{hh mm ss.sss} & \colhead{$\degr~~\arcmin~~\arcsec$} & \colhead{}
}
\startdata 
SMC\,3\_3 &    00 44 55.896 &$-$74 12 42.120 &    315           \\
SMC\,3\_5 &    01 27 30.816 &$-$74 00 49.320 &    25             \\
SMC\,4\_2 &    00 25 14.088 &$-$73 01 47.640 &    86             \\
SMC\,4\_3 &    00 45 14.688 &$-$73 07 11.280 &    1642             \\
SMC\,4\_4 &    01 05 19.272 &$-$73 05 15.360 &    1128             \\
SMC\,4\_5 &    01 25 11.088 &$-$72 56 02.760 &    83             \\
SMC\,5\_2 &    00 26 41.688 &$-$71 56 35.880 &    2             \\
SMC\,5\_3 &    00 45 32.232 &$-$72 01 40.080 &    197             \\
SMC\,5\_4 &    01 04 26.112 &$-$71 59 51.000 &    687             \\
SMC\,6\_3 &    00 45 48.792 &$-$70 56 09.240 &    4             \\
SMC\,6\_5 &    01 21 22.560 &$-$70 46 11.640 &    3             \\
\enddata 
\end{deluxetable}

\begin{deluxetable}{ccccccc}
\tabletypesize{\scriptsize}
\tablecaption{Number of CCs for each different mode of pulsation.\label{ncep2}}
\tablewidth{0pt}
\tablehead{
\colhead{F} & \colhead{FO} & \colhead{SO} & \colhead{F/FO} & \colhead{FO/SO} & \colhead{F/FO/SO} & \colhead{FO/SO/TO} 
}
\startdata 
2377 &  1472 & 74 & 52 & 196 & 2& 1 \\ 
\enddata 
\end{deluxetable}

\begin{deluxetable}{ccc}
\tabletypesize{\scriptsize}
\tablecaption{$Y$, $J$ and $K_\mathrm{s}$ time series photometry for 
  the CCs investigated in this paper. The data below refer to the variable 
OGLE-SMC-CEP-2476. \label{VMCPhot}}
\tablewidth{0pt}
\tablehead{
\colhead{$HJD-2 400 000$} & \colhead{$Y$} & \colhead{$\sigma_Y$} 
}
\startdata 
   55492.59731 &  19.076 &   0.040  \\
   55492.63328 &  19.130 &   0.040  \\
   55497.70319 &  19.116 &   0.047  \\
   55539.61969 &  19.114 &   0.051  \\
\cutinhead{$HJD-2 400 000$ ~~~~~~~  $J$ ~~~~~~~~~ $\sigma_J$ ~~~} 
   55493.58975 &  18.788 &   0.041  \\
   55493.62870 &  18.830 &   0.039  \\
   55495.55177 &  18.860 &   0.063  \\
   55539.64087 &  18.912 &   0.060  \\
   55778.75378 &  18.817 &   0.052  \\
\cutinhead{$HJD-2 400 000$ ~~~~~~~  $K_\mathrm{s}$ ~~~~~~~~ $\sigma_{K_\mathrm{s}}$ ~~~} 
   55493.78892 &  18.641 &   0.099  \\
   55495.57575 &  18.699 &   0.168  \\
   55495.68566 &  18.688 &   0.111  \\
   55497.72461 &  18.776 &   0.153  \\
   55538.62081 &  18.681 &   0.125  \\
   55549.58532 &  18.777 &   0.135  \\
   55769.75425 &  18.697 &   0.126  \\
   55778.77517 &  18.683 &   0.169  \\
   55791.76203 &  18.706 &   0.110  \\
   55818.73171 &  18.807 &   0.148  \\
   55820.67458 &  18.484 &   0.088  \\
   55879.55553 &  18.542 &   0.111  \\
   55880.61929 &  18.737 &   0.137  \\
   55900.57090 &  18.674 &   0.115  \\
   56130.79475 &  18.704 &   0.116  \\
   56173.70295 &  18.712 &   0.136  \\
   56195.64097 &  18.668 &   0.111  \\
   56223.54177 &  18.704 &   0.125  \\
\enddata 
\tablecomments{Table \ref{VMCPhot} is published in its entirety in the 
electronic edition of the {\it Astrophysical Journal}.  A portion is 
shown here for guidance regarding its form and content.}
\end{deluxetable}

\begin{deluxetable*}{cccccccc}
\tablewidth{0pt} \tablecaption{Fourier parameters adopted to construct 
  the templates in the VISTA $Y,J,K_\mathrm{s}$ bands. Note that the 
  template starts from 2 because template 1 (T1) is a simple cosine 
  function.\label{paramFourier}} 
\tablehead{ 
\colhead{Parameter} & \colhead{T2} & \colhead{T3} & \colhead{T4} &
\colhead{T5} & \colhead{T6} & \colhead{T7} & \colhead{T8}} 
\startdata 
\cutinhead{$Y$--Band} $a_1 $ & 0.49260 & 0.12614 & 0.12888 & 0.20863 &
0.20097 & 0.17442 & 0.15502 \\ $a_2 $ & 0.14500 & 0.03315 & 0.07549 &
0.11718 & 0.09804 & 0.05488 & 0.03855 \\ $a_3 $ & 0.04100 & 0.01090 &
0.04152 & 0.05601 & 0.04747 & 0.02273 & 0.02314 \\ $a_4 $ & 0.01000 &
0.00448 & 0.01609 & 0.02200 & 0.01831 & 0.01211 & 0.00984 \\ $a_5 $ &
0.00000 & 0.00209 & 0.00580 & 0.00765 & 0.00370 & 0.00640 & 0.00483 
\\ $a_6 $ & 0.00000 & 0.00124 & 0.00000 & 0.00253 & 0.00190 & 0.00345 
& 0.00146 \\ $a_7 $ & 0.00000 & 0.00065 & 0.00000 & 0.00026 & 0.00276 
& 0.00171 & 0.00078 \\ $a_8 $ & 0.00000 & 0.00035 & 0.00000 & 0.00044 
& 0.00173 & 0.00081 & 0.00054 \\ $a_9 $ & 0.00000 & 0.00022 & 0.00000 
& 0.00008 & 0.00054 & 0.00031 & 0.00041 \\ $a_{10}$ & 0.00000 &
0.00017 & 0.00000 & 0.00027 & 0.00024 & 0.00010 & 0.00022 \\ $\phi_1$
& 1.40800 & 5.75860 & 2.52701 & 1.50221 & 0.72710 & 5.27327 & 5.93367 
\\ $\phi_2$ & 2.52800 & 3.65703 & 3.36472 & 1.50458 & 6.08919 &
3.02993 & 4.57391 \\ $\phi_3$ & 3.55200 & 1.23267 & 4.18291 & 1.34113 
& 5.11517 & 1.14020 & 3.82891 \\ $\phi_4$ & 4.43800 & 5.38378 &
5.03844 & 0.97634 & 4.16420 & 5.46810 & 1.84753 \\ $\phi_5$ & 0.00000 
& 3.36130 & 5.72830 & 0.30153 & 3.35092 & 3.52010 & 1.18378 
\\ $\phi_6$ & 0.00000 & 1.20985 & 0.00000 & 5.90041 & 5.31224 &
1.54636 & 4.81460 \\ $\phi_7$ & 0.00000 & 5.24591 & 0.00000 & 0.00420 
& 4.58734 & 5.86030 & 5.81304 \\ $\phi_8$ & 0.00000 & 3.03051 &
0.00000 & 1.40454 & 3.82023 & 3.83808 & 0.48183 \\ $\phi_9$ & 0.00000 
& 1.01212 & 0.00000 & 0.34470 & 3.26788 & 1.67678 & 3.77426 
\\ $\phi_{10}$ & 0.00000 & 5.12881 & 0.00000 & 3.91317 & 4.72011 &
5.36878 & 4.02321 \\ \cutinhead{$J$--Band} $a_1 $ & 0.49260 & 0.10773 
& 0.12934 & 0.10873 & 0.12532 & 0.12790 & 0.15502 \\ $a_2 $ & 0.14500 
& 0.01461 & 0.06480 & 0.04565 & 0.03672 & 0.05709 & 0.03855 \\ $a_3 $
& 0.04100 & 0.02933 & 0.03052 & 0.02487 & 0.00939 & 0.00983 & 0.02314 
\\ $a_4 $ & 0.01000 & 0.00196 & 0.01328 & 0.00825 & 0.00303 & 0.00215 
& 0.00984 \\ $a_5 $ & 0.00000 & 0.00177 & 0.00547 & 0.00220 & 0.00170 
& 0.00151 & 0.00483 \\ $a_6 $ & 0.00000 & 0.00010 & 0.00254 & 0.00091 
& 0.00020 & 0.00072 & 0.00146 \\ $a_7 $ & 0.00000 & 0.00013 & 0.00135 
& 0.00018 & 0.00012 & 0.00021 & 0.00078 \\ $a_8 $ & 0.00000 & 0.00011 
& 0.00061 & 0.00057 & 0.00040 & 0.00002 & 0.00054 \\ $a_9 $ & 0.00000 
& 0.00004 & 0.00023 & 0.00046 & 0.00023 & 0.00003 & 0.00041 
\\ $a_{10}$ & 0.00000 & 0.00004 & 0.00019 & 0.00031 & 0.00024 &
0.00006 & 0.00022 \\ $\phi_1$ & 1.40800 & 0.22905 & 2.72672 & 5.39342 
& 1.09258 & 5.12094 & 5.93367 \\ $\phi_2$ & 2.52800 & 4.86405 &
4.06800 & 3.37087 & 0.70885 & 2.95240 & 4.57391 \\ $\phi_3$ & 3.55200 
& 3.12523 & 5.42088 & 1.47564 & 0.70180 & 0.78702 & 3.82891 
\\ $\phi_4$ & 4.43800 & 1.79845 & 0.59352 & 5.60246 & 1.22024 &
5.53095 & 1.84753 \\ $\phi_5$ & 0.00000 & 1.00316 & 2.17713 & 2.89183 
& 1.40527 & 2.77300 & 1.18378 \\ $\phi_6$ & 0.00000 & 5.97019 &
3.76033 & 0.28963 & 0.69902 & 0.39488 & 4.81460 \\ $\phi_7$ & 0.00000 
& 5.69884 & 5.19037 & 0.47083 & 5.83421 & 5.20192 & 5.81304 
\\ $\phi_8$ & 0.00000 & 5.29476 & 0.22745 & 5.03774 & 5.32133 &
3.10882 & 0.48183 \\ $\phi_9$ & 0.00000 & 2.24848 & 1.14384 & 2.56761 
& 5.91469 & 2.69578 & 3.77426 \\ $\phi_{10}$ & 0.00000 & 2.92779 &
2.44216 & 6.20944 & 5.71899 & 5.71612 & 4.02321 
\\ \cutinhead{$K_\mathrm{s}$--Band} $a_1 $ & 0.49260 & 0.11057 &
0.18142 & 0.10520 & 0.15750 & 0.10319 & 0.10789 \\ $a_2 $ & 0.14500 &
0.04102 & 0.02969 & 0.03653 & 0.02199 & 0.03305 & 0.04682 \\ $a_3 $ &
0.04100 & 0.01533 & 0.00711 & 0.01983 & 0.00921 & 0.01029 & 0.02355 
\\ $a_4 $ & 0.01000 & 0.00440 & 0.00214 & 0.01085 & 0.00367 & 0.00155 
& 0.01431 \\ $a_5 $ & 0.00000 & 0.00089 & 0.00302 & 0.00550 & 0.00224 
& 0.00001 & 0.00800 \\ $a_6 $ & 0.00000 & 0.00078 & 0.00146 & 0.00259 
& 0.00119 & 0.00011 & 0.00448 \\ $a_7 $ & 0.00000 & 0.00048 & 0.00051 
& 0.00127 & 0.00064 & 0.00032 & 0.00232 \\ $a_8 $ & 0.00000 & 0.00004 
& 0.00069 & 0.00079 & 0.00036 & 0.00033 & 0.00088 \\ $a_9 $ & 0.00000 
& 0.00016 & 0.00016 & 0.00058 & 0.00028 & 0.00018 & 0.00028 
\\ $a_{10}$ & 0.00000 & 0.00014 & 0.00049 & 0.00045 & 0.00021 &
0.00000 & 0.00000 \\ $\phi_1 $ & 1.40800 & 1.27522 & 1.10728 & 1.31803 
& 1.57541 & 4.75055 & 1.48634 \\ $\phi_2 $ & 2.52800 & 1.80953 &
2.41492 & 2.06525 & 2.75238 & 2.11317 & 2.14552 \\ $\phi_3 $ & 3.55200 
& 2.46781 & 2.77198 & 2.84018 & 4.16989 & 5.38840 & 2.71326 \\ $\phi_4 
$ & 4.43800 & 2.94883 & 2.45223 & 3.61468 & 5.36458 & 2.83264 &
3.39001 \\ $\phi_5 $ & 0.00000 & 2.46814 & 2.58196 & 4.34756 & 0.48850 
& 4.55134 & 3.97662 \\ $\phi_6 $ & 0.00000 & 1.69940 & 3.83981 &
4.99929 & 2.10232 & 1.60734 & 4.48671 \\ $\phi_7 $ & 0.00000 & 2.00844 
& 2.89989 & 5.50476 & 3.90737 & 5.88377 & 5.06340 \\ $\phi_8 $ &
0.00000 & 2.26548 & 4.29762 & 5.96834 & 5.73100 & 3.05768 & 5.44150 
\\ $\phi_9 $ & 0.00000 & 6.16960 & 3.43667 & 0.23028 & 1.22432 &
0.08505 & 5.61231 \\ $\phi_{10}$ & 0.00000 & 0.16239 & 4.88214 &
0.80864 & 2.94561 & 4.77334 & 0.00000 \\ \enddata 
\end{deluxetable*}

\clearpage 
\begin{turnpage}

\begin{deluxetable}{@{}c@{}c@{}cccc@{}c@{}c@{}c@{}c@{}c@{}c@{}c@{}c@{}c@{}c@{}c@{}c@{}c@{}c@{}c@{}c}
\tabletypesize{\scriptsize} 
\tablecaption{Results of the template-fitting procedure. Columns: (1)  
  Identification from OGLE\,III (OGLE-SMC-CEP- plus the numbers listed  
  below) or EROS\,2; (2) $VMC$ tile in which the object is found; (3)  
  Right Ascension; (4) Declination; (5) Mode: F=Fundamental; FO=First  
  Overtone; SO=Second overtone; TO=Third Overtone; (6) Period; (7)  
  Number of epochs in $Y$; (8)--(9) Intensity-averaged magnitude in  
  $Y$ and relative uncertainty; (10)--(11) Peak-to-peak amplitude in  
  $Y$ and relative uncertainty; (12) to (16) As for column (7) to (11)  
  but for the $J$ band; (17) to (21) As for column (7) to (11) but for  
  the $K_\mathrm{s}$ band; (22) Flag assigned using the  
  $PW(V,K_\mathrm{s})$ as reference relation: 0=no remark;
  1=overluminous star with separation between $VMC$ and OGLE\,III  
  position larger than 0.2 arcsec; 2=largely scattered or  
  under-sampled light curve; 3=Remarks OGLE\,III; 4= Remarks $VMC$;
  5=Saturation; 10--11=F- or FO-mode outliers without evident  
  explanation; 12=$V$ band lacking. The first part of the table  
  includes all stars with flag = 0 ordered by increasing period; the  
  second part includes all stars with flag$>$0, sorted by increasing  
  period. The sorting of the table is the same as in  
  Figs~\ref{figY_pag119}, ~\ref{figJ_pag119}, and \ref{figK_pag119}.  
  We show the first 20 rows of the table to indicate its form and  
  content. 
\label{tabResults}}
\tablehead{
\colhead{ID} & \colhead{Tile} &
\colhead{RA} & \colhead{DEC} & \colhead{Mode} & \colhead{Period}  & 
\colhead{n$_Y$}   & \colhead{$Y$}   & \colhead{$\sigma(Y)$} & \colhead{A($Y$)}   & \colhead{$\sigma$A($Y$)}  &
\colhead{n$_J$}   & \colhead{$J$}   & \colhead{$\sigma(J)$} & \colhead{A($J$)}   & \colhead{$\sigma$A($J$)}  &
\colhead{n$_{K_\mathrm{s}}$}   & \colhead{$K_\mathrm{s}$}   &
\colhead{$\sigma(K_\mathrm{s})$} & \colhead{A($K_\mathrm{s}$)}   &
\colhead{$\sigma$A($K_\mathrm{s}$)} & \colhead{Flag} \\
\colhead{} & \colhead{} & \colhead{deg} & \colhead{deg} &
\colhead{} & \colhead{d}   &
\colhead{}   & \colhead{mag}   & \colhead{mag} & \colhead{mag}   &
\colhead{mag}  & \colhead{}   & \colhead{mag}   & \colhead{mag} & \colhead{mag}   &
\colhead{mag}  & \colhead{}   & \colhead{mag}   & \colhead{mag} & \colhead{mag}   & \colhead{mag}  & \colhead{} \\
\colhead{1} & \colhead{2} & \colhead{3} & \colhead{4} &
\colhead{5} & \colhead{6}   &
\colhead{7}   & \colhead{8}   & \colhead{9} & \colhead{10}   &
\colhead{11}  & \colhead{12}   & \colhead{13}   & \colhead{14} & \colhead{15}   &
\colhead{16}  & \colhead{17}   & \colhead{18}   & \colhead{19} & \colhead{20}   & \colhead{21}  & \colhead{22} 
}
\startdata  
2476  &  5\_4    &    13.991833  &   -72.442611 &     FO/SO  &   0.252601  &  4  &   19.058  &   0.087  &   0.10  & 0.09   &  5  &   18.873  &   0.056  &   0.15  & 0.09  & 18  &   18.696  &   0.057  &   0.110  &   0.085   &               0       \\ 
3867  &  4\_4    &    16.676208  &   -73.416917 &  FO/SO/TO  &   0.268847  &  4  &   18.454  &   0.026  &   0.19  & 0.04   &  5  &   18.245  &   0.029  &   0.15  & 0.06  & 14  &   18.041  &   0.036  &   0.050  &   0.052   &               0       \\ 
2507  &  4\_4    &    14.038583  &   -73.251389 &     FO/SO  &   0.277552  &  4  &   18.645  &   0.026  &   0.11  & 0.05   &  5  &   18.422  &   0.004  &   0.10  & 0.01  & 14  &   18.213  &   0.065  &   0.184  &   0.092   &               0       \\ 
0022  &  4\_2    &     5.892208  &   -73.399694 &        FO  &   0.313664  &  5  &   18.442  &   0.004  &   0.17  & 0.01   &  5  &   18.268  &   0.021  &   0.10  & 0.03  & 14  &   18.061  &   0.035  &   0.087  &   0.064   &               0       \\ 
1471  &  5\_3    &    12.571750  &   -72.043333 &     FO/SO  &   0.327181  &  5  &   17.736  &   0.014  &   0.07  & 0.02   &  9  &   17.550  &   0.010  &   0.11  & 0.02  & 16  &   17.326  &   0.017  &   0.065  &   0.019   &               0       \\ 
3287  &  4\_4    &    15.427042  &   -73.294194 &        FO  &   0.346422  &  4  &   18.061  &   0.014  &   0.14  & 0.04   &  5  &   17.878  &   0.015  &   0.11  & 0.03  & 14  &   17.676  &   0.027  &   0.126  &   0.037   &               0       \\ 
1606  &  4\_3    &    12.773167  &   -73.377056 &     FO/SO  &   0.352508  &  6  &   18.313  &   0.021  &   0.20  & 0.05   &  6  &   18.074  &   0.004  &   0.11  & 0.01  & 16  &   17.784  &   0.031  &   0.070  &   0.035   &               0       \\ 
3784  &  4\_4    &    16.470875  &   -72.891500 &        FO  &   0.380368  &  4  &   18.322  &   0.055  &   0.15  & 0.05   &  5  &   18.009  &   0.017  &   0.18  & 0.04  & 14  &   17.765  &   0.028  &   0.087  &   0.035   &               0       \\ 
4243  &  4\_4    &    17.760458  &   -73.161833 &        FO  &   0.393727  &  4  &   18.131  &   0.013  &   0.28  & 0.03   &  5  &   17.907  &   0.010  &   0.16  & 0.03  & 14  &   17.643  &   0.036  &   0.071  &   0.054   &               0       \\ 
0310  &  4\_3    &     9.822625  &   -73.240444 &        FO  &   0.394238  &  6  &   17.899  &   0.017  &   0.19  & 0.05   &  6  &   17.696  &   0.016  &   0.16  & 0.03  & 16  &   17.540  &   0.021  &   0.146  &   0.040   &               0       \\ 
1357  &  4\_3    &    12.404083  &   -73.025194 &        SO  &   0.401286  &  6  &   17.693  &   0.010  &   0.05  & 0.01   &  6  &   17.460  &   0.012  &   0.03  & 0.02  & 16  &   17.255  &   0.017  &   0.043  &   0.026   &               0       \\ 
2265  &  3\_3    &    13.668625  &   -73.801333 &        FO  &   0.408552  &  7  &   17.828  &   0.019  &   0.12  & 0.05   &  5  &   17.692  &   0.018  &   0.21  & 0.03  & 18  &   17.512  &   0.015  &   0.103  &   0.030   &               0       \\ 
4465  &  4\_5    &    18.738000  &   -72.666972 &        FO  &   0.421663  &  5  &   17.868  &   0.010  &   0.19  & 0.04   &  6  &   17.691  &   0.009  &   0.15  & 0.02  & 18  &   17.472  &   0.017  &   0.103  &   0.028   &               0       \\ 
0358  &  4\_3    &    10.121583  &   -73.447056 &        SO  &   0.442236  &  6  &   17.205  &   0.007  &   0.05  & 0.02   &  6  &   17.057  &   0.009  &   0.03  & 0.02  & 16  &   16.912  &   0.010  &   0.020  &   0.016   &               0       \\ 
2290  &  3\_3    &    13.713833  &   -73.872278 &        FO  &   0.444418  &  7  &   18.045  &   0.012  &   0.23  & 0.03   &  7  &   17.853  &   0.015  &   0.17  & 0.03  & 18  &   17.564  &   0.019  &   0.128  &   0.031   &               0       \\ 
4618  &  4\_5    &    20.858250  &   -72.729806 &        SO  &   0.445367  &  5  &   17.100  &   0.005  &   0.05  & 0.01   &  6  &   16.952  &   0.005  &   0.03  & 0.01  & 18  &   16.746  &   0.014  &   0.030  &   0.020   &               0       \\ 
1085  &  4\_3    &    12.000917  &   -72.865417 &        SO  &   0.452269  &  6  &   17.332  &   0.010  &   0.05  & 0.01   &  6  &   17.155  &   0.013  &   0.07  & 0.02  & 16  &   17.060  &   0.016  &   0.026  &   0.018   &               0       \\ 
4096  &  5\_4    &    17.286083  &   -72.275361 &        FO  &   0.487218  &  4  &   17.988  &   0.050  &   0.47  & 0.06   &  5  &   17.700  &   0.005  &   0.18  & 0.02  & 18  &   17.469  &   0.021  &   0.110  &   0.034   &               0       \\ 
2907  &  4\_4    &    14.730208  &   -73.550722 &     FO/SO  &   0.490294  &  4  &   17.607  &   0.048  &   0.22  & 0.05   &  5  &   17.540  &   0.016  &   0.21  & 0.03  & 14  &   17.274  &   0.019  &   0.080  &   0.031   &               0       \\ 
1133  &  3\_3    &    12.061292  &   -73.718278 &       SO   &   0.490791  & 13  &   17.253  &   0.006  &   0.02  & 0.01   & 13  &   17.063  &   0.007  &   0.01  & 0.01  & 34  &   16.905  &   0.009  &   0.014  &   0.010   &               0 \\      
\enddata 
\tablecomments{Table \ref{tabResults} is published in its entirety in the  
electronic edition of the {\it Astrophysical Journal}.  A portion is  
shown here for guidance regarding its form and content.}
\end{deluxetable}

\end{turnpage}
\clearpage  

\begin{deluxetable*}{lccccccc}
\tablewidth{0pt}
\tablecaption{$PL$, $PW$, and $PLC$ relations for F and FO CCs. The 
Wesenheit functions are defined in the table.\label{pl}}
\tablehead{
\colhead{Mode}      & \colhead{$\alpha$}      &
\colhead{$\sigma_{\alpha}$}          & \colhead{$\beta$}  &
\colhead{$\sigma_{\beta}$}          & \colhead{$\gamma$}    &
\colhead{$\sigma_{\gamma}$}          & \colhead{r.m.s.}}
\startdata 
\cutinhead{$Y^0$=$\alpha$ +$\beta$ log$P$}
F   log$P<0.47$ &  17.247  &   0.011  &  $-$3.413  &  0.043  & \nodata& \nodata & 0.196 \\
F log$P\ge0.47$ &  17.016  &   0.024  &  $-$2.984  &  0.030  & \nodata & \nodata & 0.197 \\
FO              &  16.605  &   0.006  &  $-$3.365  &  0.025  & \nodata & \nodata & 0.202 \\
SO              & 15.91    &  0.06     &  $-$4.06   & 0.27    &\nodata & \nodata & 0.15 \\
\cutinhead{$J^0$=$\alpha$ +$\beta$ log$P$}	      		                         
F   log$P<0.47$ &  16.978  &   0.010  &  $-$3.469  &  0.040  & \nodata & \nodata & 0.182 \\
F log$P\ge0.47$ &  16.763  &   0.021  &  $-$3.047  &  0.027  & \nodata & \nodata & 0.177 \\
FO              &  16.372  &   0.005  &  $-$3.416  &  0.023  & \nodata & \nodata & 0.185 \\
SO              & 15.73    &  0.06     &  $-$4.07   & 0.26    &\nodata & \nodata & 0.15 \\
\cutinhead{$K_\mathrm{s}^0$=$\alpha$+$\beta$ log$P$}
F log$P<0.47$ & 16.711  &   0.009  &  $-$3.578  &  0.036  & \nodata  & \nodata & 0.166 \\
F log$P\ge0.47$ & 16.513  &   0.019  &  $-$3.195  &  0.024  & \nodata & \nodata & 0.156 \\
FO              & 16.133  &   0.005  &  $-$3.544  &  0.020  & \nodata & \nodata & 0.169 \\
SO              & 15.52    &  0.06     &  $-$4.28   & 0.26    & \nodata & \nodata & 0.15 \\
\cutinhead{$W(Y,K_\mathrm{s})$=$K_\mathrm{s}-0.42\,(Y-K_\mathrm{s})$=$\alpha$   + $\beta$ log$P$}
F   log$P<0.47$ &  16.489  &   0.009  &  $-$3.660  &  0.035  & \nodata & \nodata & 0.158 \\
F log$P\ge0.47$ &  16.301  &   0.017  &  $-$3.283  &  0.022  & \nodata & \nodata & 0.145 \\
FO              &  15.933  &   0.005  &  $-$3.614  &  0.020  & \nodata & \nodata & 0.161 \\
SO              & 15.37    &  0.06     &  $-$4.29   & 0.26    & \nodata & \nodata & 0.14 \\
\cutinhead{$W(J,K_\mathrm{s})$=$K_\mathrm{s}-0.69\,(J-K_\mathrm{s})$=$\alpha$   + $\beta$ log$P$}
F   log$P<0.47$ &  16.535  &   0.009  &  $-$3.685  &  0.034  & \nodata & \nodata & 0.153 \\
F log$P\ge0.47$ &  16.343  &   0.017  &  $-$3.301  &  0.021  & \nodata & \nodata & 0.139 \\
FO              &  15.964  &   0.005  &  $-$3.618  &  0.019  & \nodata & \nodata & 0.156 \\
SO              & 15.39    &  0.06     &  $-$4.31   & 0.26    & \nodata &\nodata & 0.15 \\
\cutinhead{$W(V,K_\mathrm{s})$=$K_\mathrm{s}-0.13\,(V-K_\mathrm{s})$=$\alpha$   + $\beta$ log$P$}
F   log$P<0.47$ &  16.559  &   0.008  &  $-$3.666  &  0.033  & \nodata & \nodata & 0.147 \\
F log$P\ge0.47$ &  16.360  &   0.016  &  $-$3.265  &  0.021  & \nodata& \nodata & 0.137 \\
FO              &  15.984  &   0.004  &  $-$3.591  &  0.019  & \nodata & \nodata & 0.154 \\
SO              & 15.40    &  0.06     &  $-$4.28   & 0.26    & \nodata &\nodata & 0.15 \\
\cutinhead{$K_\mathrm{s}^0$=$\alpha$  + $\beta$  log$P$ + $\gamma$ $(Y-K_\mathrm{s})_0$}
F   log$P<0.47$ & 16.619  &   0.020 &  $-$3.608  &   0.036  &  0.17  &   0.03	&0.164  \\
F log$P\ge0.47$ & 16.239  &   0.035 &  $-$3.312  &   0.026  &  0.55  &   0.06	&0.146  \\
FO              & 15.923  &   0.023 &  $-$3.629  &   0.021  &  0.44  &   0.05	&0.163  \\
SO              & 15.62  &   0.10 &  -4.21  &   0.28  & $-$0.25  &   0.19	&0.16  \\
\cutinhead{$K_\mathrm{s}^0$=$\alpha$  + $\beta$  log$P$ + $\gamma$ $(J-K_\mathrm{s})_0$}
F   log$P<0.47$ & 16.535  &   0.022 &  $-$3.649  &   0.036  &  0.66  &   0.07	&0.161  \\
F log$P\ge0.47$ & 16.227  &   0.032 &  $-$3.372  &   0.028  &  1.16  &   0.11	&0.144  \\
FO              & 15.911  &   0.021 &  $-$3.657  &   0.021  &  0.92  &   0.08	&0.162  \\
SO              & 15.64  &   0.08 &  -4.23  &   0.27  & $-$0.65  &   0.30	&0.16 \\
\cutinhead{$K_\mathrm{s}^0$=$\alpha$  + $\beta$  log$P$ + $\gamma$ $(V-K_\mathrm{s})_0$}
F   log$P<0.47$ & 16.164  &   0.031 &  $-$3.776  &   0.033  &  0.445  &   0.024	&0.145  \\
F log$P\ge0.47$ & 15.879  &   0.035 &  $-$3.498  &   0.024  &  0.543  &   0.027	&0.121  \\
FO              & 15.676  &   0.025 &  -3.710  &   0.020  &  0.402  &   0.022	&0.149  \\
SO              & 15.39  &   0.12 &  -4.28  &   0.26  &  0.14  &   0.12	&0.15  \\
\enddata 
\end{deluxetable*}

\end{document}